\documentclass[12pt]{article}
\usepackage{amsmath,amsthm}

\begin{document}

\begin{center}
{\Large{\bf The osp(1,2)--covariant Lagrangian quantization
\\
\medskip
of reducible massive gauge theories}}
\\
\bigskip\bigskip
{\large{\sc B. Geyer}}
\\
\smallskip
{\it Universit\"at Leipzig, Naturwissenschaftlich--Theoretisches Zentrum}
\\
{\it 04109 Leipzig, Germany}
\\
\bigskip
{\large{\sc P.M. Lavrov}}
\\
\smallskip
{\it Tomsk State Pedagogial University, Tomsk 634041, Russia}
\\
\bigskip
{\large{\sc D. M\"ulsch}}
\\
\smallskip
{\it Wissenschaftszentrum Leipzig e.V., Leipzig 04103, Germany}
\\
\bigskip\medskip
{\small{\bf Abstract}}
\\
\end{center}

\begin{quotation}
\noindent {\small{The $osp(1,2)$--covariant Lagrangian quantization of 
irreducible gauge theories [1] is generalized to $L$--stage reducible
theories. The dependence of the generating functional of Green's functions 
on the choice of gauge in the massive case is discussed and Ward identities 
related to $osp(1,2)$ symmetry are given. Massive first--stage theories with 
closed gauge algebra are studied in detail. The generalization of the 
Chapline--Manton model and topological Yang--Mills theory to the case of 
massive fields is considered as examples.}}
\end{quotation}


\setlength{\baselineskip}{0.7cm}

\bigskip\medskip
\begin{flushleft}
{\large{\bf I. INTRODUCTION}}
\end{flushleft}
\bigskip
In a previous paper \cite{1}, a generalization of the $Sp(2)$--covariant 
Lagrangian quantization for irreducible (or zero--stage) general gauge
theories \cite{2,3,4} has been proposed which is based on the orthosymplectic 
algebra $osp(1,2)$. Within this approach it is possible to consider 
${\it massive}$ fields thus avoiding infrared divergencies otherwise occuring 
within the renormalization procedure. Moreover, this approach ensures 
symplectic invariance to all orders of perturbation theory. This is due to 
the fact that for nonvanishing mass $m$ the quantum action $S_m$ (and the 
related gauge fixed action $S_{m, {\rm ext}}$) is required to satisfy the 
generating equations of $Sp(2)$--symmetry in addition to the $m$--extended 
quantum master equations generating the extended BRST symmetry.  

The aim of the present paper is to extent this formalism to
$L$--stage reducible gauge theories, i.e. to theories having a redundant
set of linearly dependent gauge generators. In principle, every such theory
permits to single out a basis of linearly independent generators but then, 
in general, either locality or manifest relativistic covariance will be lost. 

The paper is organized as follows. In Section II we shortly review the basic 
definitions concerning the reducibility properties of the theory. 
The extended configuration space of $L$--stage reducible gauge theories is 
introduced and the $osp(1,2)$--covariant quantization procedure for these
theories is formulated. To be able to express this $osp(1,2)$--algebra through 
{\it operator identities} and to have nontrivial solutions of the generating 
equations it is necessary to introduce additional sources not present in
the $Sp(2)$--covariant formulation. Furthermore, the explicit construction of 
generating differential operators fulfilling this algebra is 
outlined. As in the case of irreducible theories mass terms destroy gauge 
independence; however, this gauge dependence disappears in the limit $m = 0$. 
In Section III we consider first--stage reducible massive theories with closed 
gauge algebra, thereby extending the solution given in \cite{1}. The problem 
of how to find the full set of necessarily required (anti)ghost and 
auxiliary fields has also been tackled in Ref. \cite{5} for the massless case 
by introducing additional structure constants and postulating some new 
structure relations. But we were neither able to confirm one of these 
relations (Eq. (15) in Ref. \cite{5}) nor to prove the nilpotency of the 
corresponding extended BRST transformations. The same inaccuracy was
adopted in Ref. \cite{6}. Re--analysing that problem we 
proved that the above mentioned relation had to be generalized (see Eq. (32) 
below) in order to ensure nilpotency. As a consequence, also quartic 
(anti)ghost terms enter into the extended BRST transformations and do not 
disappear as has been claimed in Ref. \cite{5}. In Section IV as an 
application we consider the Chapline--Manton model \cite{7} as well as 
topological Yang--Mills theory \cite{8} and generalize the corresponding 
(anti)BRST transformations for the massive case.  

Throughout this paper we have used the condensed notation introduced by 
DeWitt \cite{9} and conventions adopted in Ref. \cite{1}; if not otherwise 
specified, derivatives with respect to the antifields are the (usual) left 
ones and that with respect to the fields are {\it right} ones. Left  
derivatives with respect to the fields are labeled by the subscript $L$, 
for example, $\delta_L/\delta \phi^A$ denotes the left derivative with respect 
to the fields $\phi^A$.
\bigskip\medskip
\begin{flushleft}
{\large{\bf II. GENERAL STRUCTURE OF osp(1,2)--COVARIANT
\\
QUANTIZATION OF REDUCIBLE GAUGE THEORIES}}
\end{flushleft}
\bigskip
In general gauge theories a set of gauge (as well as matter) fields $A^i$ with 
Grassmann parity $\epsilon(A^i) = \epsilon_i$ is considered for which the 
classical action $S_{\rm cl}(A)$ is invariant under the gauge transformations 
\begin{equation}
\delta A^i = R^i_{\alpha_0} \xi^{\alpha_0},
\qquad
\alpha_0 = 1, \ldots, n_0,
\qquad
S_{{\rm cl}, i} R^i_{\alpha_0} = 0,
\end{equation}
where $\xi^{\alpha_0}$ are the parameters of these transformations and 
$R^i_{\alpha_0}(A)$ are the gauge generators having Grassmann parity 
$\epsilon(\xi^{\alpha_0}) = \epsilon_{\alpha_0}$ and  
$\epsilon(R^i_{\alpha_0}) = \epsilon_i + \epsilon_{\alpha_0}$, respectively;
by definition $X_{, j} = \delta X/ \delta A^j$.

For {\it general gauge theories} the (open) algebra of generators has the 
form \cite{2}:
\begin{equation}
R_{\alpha_0, j}^i R_{\beta_0}^j -
(-1)^{\epsilon_{\alpha_0} \epsilon_{\beta_0}} 
R_{\beta_0, j}^i R_{\alpha_0}^j =
- R_{\gamma_0}^i F_{\alpha_0 \beta_0}^{\gamma_0} - 
M^{ij}_{\alpha_0 \beta_0} S_{{\rm cl}, j},
\end{equation}
where $F_{\alpha_0 \beta_0}^{\gamma_0}(A)$ are the field dependent structure 
functions and $M^{ij}_{\alpha_0 \beta_0}(A)$ is graded antisymmetric with
respect to $(ij)$ and $(\alpha_0 \beta_0)$. In the case 
$M_{\alpha_0 \beta_0}^{ij} = 0$ the algebra is closed.

If the set of generators $R^i_{\alpha_0}$ are linearly {\it independent} then 
the theory is {\it irreducible} \cite{10}. The $Sp(2)$-- and 
$osp(1,2)$--covariant quantization of these theories have been considered 
in Ref. \cite{2,1}. If the generators $R^i_{\alpha_0}$ are 
{\it linearly dependent} then, according to the following characterization, 
the theory under consideration is called $L$--stage {\it reducible} 
\cite{11,3}: There exists a chain of field dependent on--shell zero--modes 
$Z^{\alpha_s - 1}_{\alpha_s}(A)$,
\begin{alignat*}{2}
R^i_{\alpha_0} Z^{\alpha_0}_{\alpha_1} &= 
S_{{\rm cl}, j} K^{ji}_{\alpha_1},
&\qquad
K^{ij}_{\alpha_1} &= - (-1)^{\epsilon_i \epsilon_j} K^{ji}_{\alpha_1},
\\
Z^{\alpha_{s - 2}}_{\alpha_{s - 1}} Z^{\alpha_{s - 1}}_{\alpha_s} &= 
S_{{\rm cl}, j} K^{j \alpha_{s - 2}}_{\alpha_s},
&\qquad
\alpha_s &= 1, \ldots, n_s, ~ s = 2, \ldots, L,
\end{alignat*}
where the stage $L$ of reducibility is defined by the lowest value $s$ 
for which the matrix $Z^{\alpha_{L - 1}}_{\alpha_L}(A)$ is no longer 
degenerated. The $Z^{\alpha_{s - 1}}_{\alpha_s}$ are the on--shell zero 
modes for $Z^{\alpha_{s - 2}}_{\alpha_{s - 1}}$ with 
$\epsilon(Z^{\alpha_{s - 1}}_{\alpha_s}) = \epsilon_{\alpha_{s - 1}} + 
\epsilon_{\alpha_s}$, where $\epsilon_{\alpha_s}$
is the parity of the $s$--stage gauge transformation associated with the index
$\alpha_s$. In the following, if not otherwise stated, we assume $s$ to
take on the values $s = 0, \ldots, L$, thereby including also the case of
irreducible theories. 

The whole space of (anti)fields and sources together with their Grassmann 
parities (modulo 2) is characterized by the following sets 
\begin{alignat*}{2}
\phi^A &= ( A^i, B^{\alpha_s| a_1 \cdots a_s}, C^{\alpha_s| a_0 \cdots a_s} ), 
& \qquad
\epsilon(\phi^A) &\equiv \epsilon_A = 
( \epsilon_i, \epsilon_{\alpha_s} + s, \epsilon_{\alpha_s} + s + 1 ),
\\
\bar{\phi}_A &= (\bar{A}_i, 
\bar{B}_{\alpha_s| a_1 \cdots a_s}, 
\bar{C}_{\alpha_s| a_0 \cdots a_s} ), 
& \qquad
\epsilon(\bar{\phi}_A) &= \epsilon_A,
\\
\phi^*_{A a} &= ( A^*_{i a}, 
B^*_{\alpha_s a| a_1 \cdots a_s}, C^*_{\alpha_s a| a_0 \cdots a_s} ), 
& \qquad
\epsilon(\phi^*_{A a}) &= \epsilon_A + 1,
\\
\intertext{and}
\eta_A &= (D_i, 
E_{\alpha_s| a_1 \cdots a_s}, 
F_{\alpha_s| a_0 \cdots a_s} ),
& \qquad
\epsilon(\eta_A) &= \epsilon_A,
\end{alignat*}
where the pyramids of auxiliary fields $B^{\alpha_s| a_1 \cdots a_s}$ and  
(anti)ghosts $C^{\alpha_s| a_0 \cdots a_s}$ ($s = 0, \ldots, L$) are 
$Sp(2)$--tensors of rank $s$ and $s + 1$, respectively, {\it symmetric} with 
respect to the indices behind the stroke $|$; and similary for the antifields 
and sources. Of course, the totally symmetrized tensors are irreducible and 
have maximal $Sp(2)$--spin.

Raising and lowering of $Sp(2)$--indices is obtained by the invariant tensor 
of the group,
\begin{equation*}
\epsilon^{ab} = \begin{pmatrix} 0 & 1 \\ -1 & 0 \end{pmatrix},
\qquad
\epsilon^{ac} \epsilon_{cb} = \delta^a_b.
\end{equation*}
Let us point to the fact that in the $Sp(2)$--approach the internal $Sp(2)$ 
indices $a_0, \ldots, a_s$ of the component fields behind the stroke $|$ are 
dummy ones, i.e. they are not affected by main operations like antibrackets
$( ~, ~ )^a$, operators $\Delta^a$, $V^a$ being introduced there.

Let us now repeat the general modifications of the $Sp(2)$--formalism
introduced in Ref. \cite{1} to obtain the $osp(1,2)$--covariant quantization
which also apply to $L$--stage reducible theories of massive fields whose 
bosonic action $S_m = S_m(\phi^A, \phi^*_{A a}, \bar{\phi}_A, \eta_A)$ depends 
on the mass $m$ as a further independent parameter. In addition to the 
$m$--{\it extended} generalized quantum master equations which ensure 
(anti)BRST invariance, $S_m$ is required to obey the generating equations of 
$Sp(2)$--invariance, too:
\begin{equation}
\bar{\Delta}_m^a\, {\rm exp}\{ (i/ \hbar) S_m \} = 0,
\qquad
\bar{\Delta}_\alpha\, {\rm exp}\{ (i/ \hbar) S_m \} = 0,
\end{equation}
or equivalently,
\begin{equation}
\hbox{$\frac{1}{2}$} ( S_m, S_m )^a + V_m^a S_m =
i \hbar \Delta^a S_m,
\qquad
\hbox{$\frac{1}{2}$} \{ S_m, S_m \}_\alpha + V_\alpha S_m =
i \hbar \Delta_\alpha S_m;
\end{equation}
$\bar{\Delta}_m^a = \Delta^a + (i/ \hbar) V_m^a$ and 
$\bar{\Delta}_\alpha = \Delta_\alpha + (i/ \hbar) V_\alpha$ are odd and even
second--order differential operators, respectively; together with the 
brackets $( S_m, S_m )^a$ and $\{ S_m, S_m \}_\alpha$ they are defined below
Eqs. (8)--(12). 
As long as $m \neq 0$ the operators $\bar{\Delta}_m^a$ are neither nilpotent 
nor do they anticommute among themselves; instead, together with the operators 
$\bar{\Delta}_\alpha$ they form the (super)algebra $osp(1,2)$:
\begin{align}
[ \bar{\Delta}_\alpha, \bar{\Delta}_\beta ] &= (i/ \hbar)
\epsilon_{\alpha\beta}^{~~~\!\gamma} \bar{\Delta}_\gamma, 
\\ 
[ \bar{\Delta}_\alpha, \bar{\Delta}_m^a ] &= (i/ \hbar) 
\bar{\Delta}_m^b (\sigma_\alpha)_b^{~a}, 
\\ 
\{ \bar{\Delta}_m^a, \bar{\Delta}_m^b \} &= - (i/ \hbar) m^2 
(\sigma_\alpha)^{ab} \bar{\Delta}^\alpha. 
\end{align}
The matrices $\sigma_\alpha$ ($\alpha = 0,+,-$) generate $sl(2, R)$, the even 
part of $osp(1,2)$, which is isomorphic to $sp(2, R)$, 
\begin{gather*}
(\sigma_\alpha)_a^{~c} (\sigma_\beta)_c^{~b} = g_{\alpha\beta} \delta_a^b + 
\hbox{$\frac{1}{2}$} \epsilon_{\alpha\beta\gamma} (\sigma^\gamma)_a^{~b},
\qquad
(\sigma^\alpha)_a^{~b} = g^{\alpha\beta} (\sigma_\beta)_a^{~b},
\\
g^{\alpha\beta} = \begin{pmatrix} 1 & 0 & 0 \\  0 & 0 & 2 \\ 0 & 2 & 0 
\end{pmatrix},
\qquad
g^{\alpha\gamma} g_{\gamma\beta} = \delta^\alpha_\beta,
\end{gather*}
and are expressed through the Pauli matrices $\tau_\alpha$ ($\alpha = 1,2,3$) 
as $(\sigma_0)_a^{~b} = (\tau_3)_a^{~b}$, 
$(\sigma_\pm)_a^{~b} = - \hbox{$\frac{1}{2}$} (\tau_1 \pm i \tau_2)_a^{~b}$. 
Here, $\epsilon_{\alpha\beta\gamma}$ is the antisymmetric tensor, 
$\epsilon_{0+-} = 1$. As has been pointed out in Ref. \cite{1} the ghost 
number operator is $(\hbar/ i) \bar{\Delta}_0 = \Delta_{gh}$.
 
In writing Eqs. (4) we have introduced the (anti)brackets $( F,G )^a$ and
$\{ F,G \}_\alpha$ defining the well known odd graded and a new even graded algebraic 
structure on the space of fields and antifields, respectively, 
\begin{align}
( F,G )^a &= 
\frac{\delta F}{\delta \phi^A}
\frac{\delta G}{\delta \phi^*_{A a}} - 
(-1)^{(\epsilon(F) + 1) (\epsilon(G) + 1)} 
( F \leftrightarrow G ),
\\
\{ F,G \}_\alpha &= (\sigma_\alpha)_B^{~~\!A}  
\frac{\delta F}{\delta \phi^A} \frac{\delta G}{\delta \eta_B} +
(-1)^{\epsilon(F) \epsilon(G)} 
( F \leftrightarrow G ),
\end{align}
whose properties were analyzed in Ref. \cite{1}. The first--order differential 
operators $V_m^a$ and $V_\alpha$ are given by
\begin{align}
V_m^a &= 
\epsilon^{ab} \phi^*_{A b} \frac{\delta}{\delta \bar{\phi}_A} -
\eta_A \frac{\delta}{\delta \phi^*_{A a}} +
m^2 (P_+)^{B a}_{A b} \bar{\phi}_B 
\frac{\delta}{\delta \phi^*_{A b}} -
m^2 \epsilon^{ab} (P_-)^{B c}_{A b} \phi^*_{B c} 
\frac{\delta}{\delta \eta_A},
\\
V_\alpha &=   
\bar{\phi}_B (\sigma_\alpha)^B_{~~\!A} 
\frac{\delta}{\delta \bar{\phi}_A} + \bigr( 
\phi^*_{A b} (\sigma_\alpha)^b_{~a} +
\phi^*_{B a} (\sigma_\alpha)^B_{~~\!A} \bigr)
\frac{\delta}{\delta \phi^*_{A a}} + 
\eta_B (\sigma_\alpha)^B_{~~\!A} 
\frac{\delta}{\delta \eta_A}, 
\end{align}
and the second--order differential operators $\Delta^a$ and $\Delta_\alpha$, 
whose structure is extracted from (8) and (9), are
\begin{equation}
\Delta^a = (-1)^{\epsilon_A}
\frac{\delta_L}{\delta \phi^A} \frac{\delta}{\delta \phi^*_{A a}}, 
\qquad
\Delta_\alpha = (-1)^{\epsilon_A} (\sigma_\alpha)_B^{~~\!A}
\frac{\delta_L}{\delta \phi^A} \frac{\delta}{\delta \eta_B}.
\end{equation}
As in \cite{1} the strategy to define the operators 
$\bar{\Delta}_m^a = \Delta^a + (i/ \hbar) V_m^a$, 
$\bar{\Delta}_\alpha = \Delta_\alpha + (i/ \hbar) V_\alpha$ is governed by a 
specific realization of the (anti)BRST-- and $Sp(2)$--transformations of the 
{\it antifields}. In accordance with (10) and (11) the action of $V_m^a$ and
$V_\alpha$ on the antifields is given by
\begin{alignat}{2}
V_m^a \bar{\phi}_A &= \epsilon^{ab} \phi^*_{A b},
& \qquad
V_\alpha \bar{\phi}_A &= 
\bar{\phi}_B (\sigma_\alpha)^B_{~A},
\nonumber\\
V_m^a \phi_{A b}^* &= m^2 (P_+)^{B a}_{A b} \bar{\phi}_B - 
\delta^a_b \eta_A, 
& \qquad
V_\alpha \phi^*_{A a} &=
\phi^*_{A b} (\sigma_\alpha)^b_{~a} +
\phi^*_{B a} (\sigma_\alpha)^B_{~A},
\\
V_m^a \eta_A &= - m^2 \epsilon^{ab} (P_-)^{B c}_{A b} \phi^*_{B c},
& \qquad
V_\alpha \eta_A &=
\eta_B (\sigma_\alpha)^B_{~A},
\nonumber
\end{alignat}
where the following abbreviations are used:
\begin{equation*}
(P_-)^{B a}_{A b} \equiv (P_+)^{B a}_{A b} - (P_+)^B_A \delta^a_b +
\delta^B_A \delta^a_b,
\qquad
(P_+)^B_A \equiv \delta^b_a (P_+)^{B a}_{A b},
\qquad
(\sigma_\alpha)^B_{~A} \equiv (\sigma_\alpha)^b_{~a} (P_+)^{B a}_{A b}.
\end{equation*}
The transformations (13) have the same form as in the irreducible case
except for the matrix $(P_+)^{B a}_{A b}$ which obviously has to be 
generalized as follows:
\begin{equation*}
(P_+)^{B a}_{A b} \equiv \begin{cases} 
\delta^i_j \delta^a_b
& \text{for $A = i, B = j$},
\\
\delta^{\beta_s}_{\alpha_s} (s + 1) 
S^{b_1 \cdots b_s a}_{a_1 \cdots a_s b}  
& \text{for $A = \alpha_s|a_1 \cdots a_s, B = \beta_s|b_1 \cdots b_s$},
\\
\delta^{\beta_s}_{\alpha_s} (s + 2) 
S^{b_0 \cdots b_s a}_{a_0 \cdots a_s b}  
& \text{for $A = \alpha_s|a_0 \cdots a_s, B = \beta_s|b_0 \cdots b_s$},
\\
0 & \text{otherwise}
\end{cases}
\end{equation*}
where the symmetrizer $S^{b_0 \cdots b_s a}_{a_0 \cdots a_s b}$ is 
defined as
\begin{equation*}
S^{b_0 \cdots b_s a}_{a_0 \cdots a_s b} \equiv 
\frac{1}{(s + 2)!} \frac{\partial}{\partial X^{a_0}} \cdots
\frac{\partial}{\partial X^{a_s}} \frac{\partial}{\partial X^b} 
X^a X^{b_s} \cdots X^{b_0},
\end{equation*}
so that $S^{b_0 \cdots b_s a}_{c_0 \cdots c_s d} 
S^{c_0 \cdots c_s d}_{a_0 \cdots a_s b} =
S^{b_0 \cdots b_s a}_{a_0 \cdots a_s b}$, $X^a$ being independent bosonic 
variables; it possesses the properties
\begin{align*}
S^{b_0 \cdots b_s a}_{a_0 \cdots a_s b} &= \frac{1}{s + 2} \bigr(
\sum_{r = 0}^s \delta^{b_r}_{a_0} 
S^{b_0 \cdots b_{r - 1} b_{r + 1} \cdots b_s a}_{a_1 \cdots a_s b} +
\frac{1}{s + 1} \sum_{r = 0}^s \delta^a_{a_0} \delta^{b_r}_b
S^{b_0 \cdots b_{r - 1} b_{r + 1} \cdots b_s}_{a_1 \cdots a_s} \bigr),
\\ 
S^{b_0 \cdots b_s}_{a_0 \cdots a_s} &= \frac{1}{s + 1}
\sum_{r = 0}^s \delta^{b_r}_{a_0}
S^{b_0 \cdots b_{r - 1} b_{r + 1} \cdots b_s}_{a_1 \cdots a_s}. 
\end{align*}

The matrices $(P_-)^B_A \equiv \delta^b_a (P_-)^{B a}_{A b}$ and 
$(\sigma_\alpha)^B_{~A}$ act nontrivially on the components of the 
(anti)fields having (dummy) internal $Sp(2)$ indices. For example,
\begin{align*}
(P_-)^B_A \bar{\phi}_B &= ( 0,
- s \bar{B}_{\alpha_s|a_1 \cdots a_s},
- (s + 1) \bar{C}_{\alpha_s|a_0 \cdots a_s} ), 
\\
\bar{\phi}_B (\sigma_\alpha)^B_{~A} &= ( 0,
\sum_{r = 1}^s \bar{B}_{\alpha_s| a_1 \cdots a_{r - 1} b a_{r + 1} \cdots a_s}
(\sigma_\alpha)^b_{~a_r},
\sum_{r = 0}^s \bar{C}_{\alpha_s|a_0 \cdots a_{r - 1} b a_{r + 1} \cdots a_s}
(\sigma_\alpha)^b_{~a_r} ).
\end{align*}
Therefore, $V_\alpha$ acts only on the (anti)ghost part of the antifields, 
and $V_m^a$ is partly of that kind (a componentwise notation of the 
transformations (13) is given in Appendix B). 

In order to prove that the transformations (13) obey the 
$osp(1,2)$--superalgebra 
\begin{equation*}
[ V_\alpha, V_\beta ] = \epsilon_{\alpha\beta}^{~~~\!\gamma} V_\gamma,
\qquad
[ V_\alpha, V_m^a ] = V_m^b (\sigma_\alpha)_b^{~a},
\qquad
\{ V_m^a, V_m^b \} = - m^2 (\sigma_\alpha)^{ab} V^\alpha
\end{equation*}
one needs the following two equalities:
\begin{align*}
\epsilon^{ad} (P_+)^{B b}_{A d} +
\epsilon^{bd} (P_+)^{B a}_{A d} &=
- (\sigma_\alpha)^{ab} (\sigma_\alpha)^d_{~c} (P_+)^{B c}_{A d},
\\
\epsilon^{ad} (P_+)^{B b}_{A c} +
\epsilon^{bd} (P_+)^{B a}_{A c} -
(\sigma_\alpha)^{ab} (\sigma_\alpha)^e_{~c} (P_-)^{B d}_{A e} &=
- (\sigma_\alpha)^{ab} \bigr(
(\sigma_\alpha)^d_{~c} \delta^B_A +
\delta^d_c (\sigma_\alpha)^B_{~A} \bigr),
\end{align*}
and the relation $(P_-)^{A b}_{C d} (P_+)^{C d}_{B a} = 0$ 
(remember that for $A = \alpha_s|a_0 \cdots a_s$, $B = \beta_s|b_0 \cdots b_s$ 
the indices $a_0 \cdots a_s$, $b_0 \cdots b_s$ are completely {\it symmetric}).
The first one is equivalent to
\begin{equation}
\epsilon^{ad} \delta^b_c + \epsilon^{bd} \delta^a_c =
- (\sigma^\alpha)^{ab} (\sigma_\alpha)^d_{~c},
\end{equation}
whereas the second one equals 
\begin{align}
& \sum_{r = 0}^s
\delta^{b_0}_{a_0} \cdots \delta^{b_{r - 1}}_{a_{r - 1}} \bigr(
\epsilon^{ad} ( \delta^b_{a_r} \delta^{b_r}_c +
\delta^b_c \delta^{b_r}_{a_r} ) +
\epsilon^{bd} ( \delta^a_{a_r} \delta^{b_r}_c +
\delta^a_c \delta^{b_r}_{a_r} ) \bigr)
\delta^{b_{r + 1}}_{a_{r + 1}} \cdots \delta^{b_s}_{a_s} 
\nonumber\\
& + (\sigma_\alpha)^{ab} \sum_{r = 0}^s
\delta^{b_0}_{a_0} \cdots \delta^{b_{r - 1}}_{a_{r - 1}} \bigr(
(s + 1) (\sigma_\alpha)^d_{~c} \delta^{b_r}_{a_r} -
(\sigma_\alpha)^{b_r}_{~~\!c} \delta^d_{a_r} ) \bigr) 
\delta^{b_{r + 1}}_{a_{r + 1}} \cdots \delta^{b_s}_{a_s} 
\\
& = - (\sigma^\alpha)^{ab} \bigr(
(\sigma_\alpha)^d_{~c}
\delta^{b_0}_{a_0} \cdots \delta^{b_s}_{a_s} +
\delta^d_c \sum_{r = 0}^s
\delta^{b_0}_{a_0} \cdots \delta^{b_{r - 1}}_{a_{r - 1}}
(\sigma_\alpha)^{b_r}_{~~\!a_r}
\delta^{b_{r + 1}}_{a_{r + 1}} \cdots \delta^{b_s}_{a_s} \bigr).
\nonumber
\end{align}
It is easily proven that every of the equalities (15) is satisfied for
$s = 1, \ldots, L$, provided the same is true for $s = 0$. Indeed,
by virtue of (14), the equations for the reducible case can be cast into 
the form 
\begin{align*}
& (\sigma^\alpha)^{ab} \sum_{r = 0}^s
\delta^{b_0}_{a_0} \cdots \delta^{b_{r - 1}}_{a_{r - 1}} \bigr(
(\sigma_\alpha)^{b_r}_{~~\!c} \delta^d_{a_r} +
\delta^{b_r}_c (\sigma_\alpha)^d_{~a_r} \bigr)
\delta^{b_{r + 1}}_{a_{r + 1}} \cdots \delta^{b_s}_{a_s} 
\\
& = (\sigma^\alpha)^{ab} \bigr(
(s + 1) (\sigma_\alpha)^d_{~c}
\delta^{b_0}_{a_0} \cdots \delta^{b_s}_{a_s} +
\delta^d_c \sum_{r = 0}^s
\delta^{b_0}_{a_0} \cdots \delta^{b_{r - 1}}_{a_{r - 1}}
(\sigma_\alpha)^{b_r}_{~~\!a_r}
\delta^{b_{r + 1}}_{a_{r + 1}} \cdots \delta^{b_s}_{a_s} \bigr)
\end{align*}
and reduce to the one for the irreducible case \cite{1},
\begin{equation}
(\sigma^\alpha)^{ab} \bigr(
(\sigma_\alpha)^{b_0}_{~~\!c} \delta^d_{a_0} +
\delta^{b_0}_c (\sigma_\alpha)^d_{~a_0} \bigr) = (\sigma^\alpha)^{ab} \bigr(
(\sigma_\alpha)^d_{~c} \delta^{b_0}_{a_0} +
\delta^d_c (\sigma_\alpha)^{b_0}_{~~\!a_0} \bigr).
\end{equation} 
The last relation (16) can be established by means of the following two
equalities:
\begin{equation*}
\epsilon^{ab} \delta^c_d + 
\epsilon^{bc} \delta^a_d + 
\epsilon^{ca} \delta^b_d = 0,
\qquad
\epsilon^{ab} ( \delta^c_e \delta^d_f - \delta^d_e \delta^c_f ) =
\epsilon^{cd} ( \delta^a_e \delta^b_f - \delta^b_e \delta^a_f ).
\end{equation*}
Let us recall that the relations (14)--(16) hold for matrices 
$\sigma_\alpha$ build up from the Pauli ones $\tau_\alpha$, ($\alpha = 1,2,3$)
($(\sigma_0)_a^{~b} = (\tau_3)_a^{~b}$, 
$(\sigma_\pm)_a^{~b} = - \hbox{$\frac{1}{2}$} (\tau_1 \pm i \tau_2)_a^{~b}$). 
In this way all definitions of Ref. \cite{1} are generalized to $L$--stage 
reducible gauge theories. Thus, the general results established in 
Ref. \cite{1} remain valid also in this case. 

The quantum action $S_m$, being a solution of Eqs. (3), (4) with the boundary 
condition $S_m|_{\phi^*_a = \bar{\phi} = \eta = \hbar = 0} = S_{\rm cl}(A)$, 
suffers from the gauge degeneracy. To remove this degeneracy an 
$Sp(2)$--{\it invariant}, gauge--fixing bosonic functional $F = F(\phi^A)$ 
has to be introduced such that the gauge fixed action $S_{m, {\rm ext}} = 
S_{m, {\rm ext}}(\phi^A, \phi^*_{A a}, \bar{\phi}_A, \eta_A)$ satisfies
Eqs. (3), (4) as well. As has been shown in Ref. \cite{1}, it is defined by
\begin{gather*}
{\rm exp}\{ (i/ \hbar) S_{m, {\rm ext}} \} = 
\hat{U}_m(F) \,{\rm exp}\{ (i/ \hbar) S_m \}, 
\\
\hat{U}_m(F) = {\rm exp}\bigr\{
\frac{\delta F}{\delta \phi^A} (
\frac{\delta}{\delta \bar{\phi}_A} -
\hbox{$\frac{1}{2}$} m^2 (P_-)^A_B \frac{\delta}{\delta \eta_B} ) -
(\hbar/ i) \hbox{$\frac{1}{2}$} \epsilon_{ab} 
\frac{\delta}{\delta \phi^*_{A a}}
\frac{\delta^2 F}{\delta \phi^A \delta \phi^B}
\frac{\delta}{\delta \phi^*_{B b}} + (i/ \hbar) m^2 F \bigr\}.
\end{gather*}
where the $\eta$--dependence of $S_m$ is restricted by the (first) condition
\begin{equation}
(\sigma_\alpha)_B^{~~\!A} \frac{\delta F}{\delta \phi^A}
\frac{\delta S_m}{\delta \eta_B} = 0,
\qquad
(\sigma_\alpha)_B^{~~\!A} \frac{\delta F}{\delta \phi^A} \phi^B = 0,
\end{equation}
such that $[ \bar{\Delta}_m^a, \hat{U}_m(F) ] 
{\rm exp}\{ (i/ \hbar) S_m \} = 0$ and
$[ \bar{\Delta}_\alpha, \hat{U}_m(F) ] {\rm exp}\{ (i/ \hbar) S_m \} = 0$.
The second condition in (17) reveals the $Sp(2)$--invariance of $F$. One of
the natural solution of the conditions (17) is
\begin{equation*}
\frac{\delta S_m}{\delta \eta_A} = \phi^A,
\end{equation*}
i.e. $S_m$ is restricted to be {\it linear} in $\eta_A$ (see Ref. \cite{1}). 
Then, as a consequence of that restriction and the tracelessness of
$\sigma_\alpha$, the second equation (4) simplifies into
\begin{equation}
(\sigma_\alpha)_B^{~~\!A} \frac{\delta S_m}{\delta \phi^A} \phi^B +
V_\alpha S_m = 0.
\end{equation}

Furthermore, let us introduce the operator
\begin{equation*}
\hat{U}_m(Y) = {\rm exp}\{(\hbar/ i) \hat{T}_m(Y)\},
\qquad
\hat{T}_m(Y) = \hbox{$\frac{1}{2}$} \epsilon_{ab} 
\{ \bar{\Delta}_m^b, [ \bar{\Delta}_m^a, Y ] \} + (i/ \hbar)^2 m^2 Y,
\end{equation*}
with $Y = Y(\phi^A, \bar{\phi}_A, \phi_{A a}^*)$ being an arbitrary (local) 
bosonic $Sp(2)$--scalar {\it independent} on $\eta_A$. Then, the operator
$\hat{U}_m(Y)$ converts any (local) solution $S_m$ of Eqs. (3) into another 
(local) solution $\tilde{S}_m$,
\begin{equation*}
{\rm exp}\{ (i/ \hbar) \tilde{S}_m \} = 
\hat{U}_m(Y) \,{\rm exp}\{ (i/ \hbar) S_m \}, 
\end{equation*}
provided it holds \cite{1}
\begin{equation*}
\frac{\delta S_m}{\delta \eta_A} = \phi^A,
\qquad
\frac{\delta Y}{\delta \eta_A} = 0,
\qquad
(\sigma_\alpha)_B^{~~\!A} \frac{\delta Y}{\delta \phi^A} \phi^B +
V_\alpha Y = 0.
\end{equation*}
Thus, the gauge itself is realized through the use of a special
transformation of this kind, namely by the operator $\hat{U}_m(Y)$ with
the special choice of $Y$ in the form $Y = F(\phi^A)$. 
 
By the same way as in Ref. \cite{1} it can be proven that the vacuum 
functional 
\begin{equation}
Z_m(0) = \int d \phi^A \,{\exp}\{(i/ \hbar) S_{m, {\rm eff}}\},
\end{equation} 
where $S_{m, {\rm eff}}(\phi^A) = 
S_{m, {\rm ext}}(\phi^A, \phi^*_{A a}, \bar{\phi}_A, \eta_A)
|_{\phi^*_a = \bar{\phi} = \eta = 0}$, is {\it not} independent on the choice 
of the gauge--fixing functional $F$ since the mass term $m^2 F$ in the action 
$S_{m, {\rm eff}}$ violates its gauge independence. However, this gauge 
dependence disappears in the limit $m = 0$ (the same is true for the 
$S$--matrix). By introducing the auxiliary fields $\pi^{A a}$, $\lambda^A$ 
and $\zeta^A$ the functional (19) can be represented in the form \cite{1}
\begin{equation}
Z_m(0) = \int d \phi^A\, d \eta_A\, d \zeta^A\, d \phi_{A a}^*\, 
d \pi^{A a}\, d \bar{\phi}_A\, d \lambda^A\,
\exp\{ (i/ \hbar) ( S_m^\zeta + W_F^\zeta - W_X ) \},
\end{equation}
with
\begin{align*}
W_F &= - \frac{\delta F}{\delta \phi^A} (
\lambda^A + \hbox{$\frac{1}{2}$} m^2 (P_+)^A_B \bar{\phi}_B ) -
\hbox{$\frac{1}{2}$} \epsilon_{ab} 
\pi^{A a} \frac{\delta^2 F}{\delta \phi^A \delta \phi^B} \pi^{B b} + m^2 F,
\\
W_X &= ( \eta_A - \hbox{$\frac{1}{2}$} m^2 (P_+)^B_A \bar{\phi}_B ) \phi^A -
\phi_{A a}^* \pi^{A a} -
\bar{\phi}_A ( \lambda^A - \hbox{$\frac{1}{2}$} m^2 (P_-)^A_B \phi^B ), 
\end{align*}
where $S_m^\zeta$ and $W_F^\zeta$ are obtained from $S_m$ and $W_F$,
respectively, by carrying out the replacement $\phi^A \rightarrow
\phi^A + \zeta^A$.

Consider the extended generating functional of the Green's functions:
\begin{equation*}
Z_m(J_A; \phi^*_{A a}, \bar{\phi}_A, \eta_A) = \int d \phi^A\, 
{\exp}\{ (i/ \hbar) ( 
S_{m, {\rm ext}}(\phi^A, \phi^*_{A a}, \bar{\phi}_A, \eta_A) + J_A \phi^A ) \}
\end{equation*} 
the generating functional of the vertex functions as usual is defined
according to
\begin{align*}
\Gamma_m(\phi^A; \phi^*_{A a}, \bar{\phi}_A, \eta_A) &=
(\hbar/ i)\, {\rm ln} Z_m(J_A; \phi^*_{A a}, \bar{\phi}_A, \eta_A) -
J_A \phi^A,
\\
\phi^A &= (\hbar/ i)
\frac{\delta {\rm ln} Z_m(J_A; \phi^*_{A a}, \bar{\phi}_A, \eta_A)}
{\delta J_A}.
\end{align*}
As consequence of the generating equations (4) for $\Gamma_m$ one gets 
the Ward identities:
\begin{equation}
\hbox{$\frac{1}{2}$} ( \Gamma_m, \Gamma_m )^a + V^a \Gamma_m = 0,
\qquad
\hbox{$\frac{1}{2}$} \{ \Gamma_m, \Gamma_m \}_\alpha + V_\alpha \Gamma_m = 0. 
\end{equation}
Moreover, if $S_m$ is restricted to be linear in $\eta_A$, then $\Gamma_m$ 
possesses the same property. In this case, according to Eq. (18), the second 
identity (21) simplifies into
\begin{equation*}
(\sigma_\alpha)_B^{~~\!A} \frac{\delta \Gamma_m}{\delta \phi^A} \phi^B +
V_\alpha \Gamma_m = 0,
\qquad
\frac{\delta \Gamma_m}{\delta \eta_A} = \phi^A.
\end{equation*}

This finishes the general introduction of the $osp(1,2)$ covariant approach 
of quantizing $L$--stage reducible general gauge theories.

\bigskip\medskip
\begin{flushleft}
{\large{\bf III. MASSIVE FIRST--STAGE REDUCIBLE THEORIES
\\
WITH A CLOSED GAUGE ALGEBRA}} 
\end{flushleft}
\bigskip
To illustrate the generalized $osp(1,2)$--quantization rules, we consider 
first--stage reducible massive theories with closed algebra. Such theories 
are characterized by the fact, first, that because 
$M_{\alpha_0 \beta_0}^{ij} = 0$, the algebra of generators, Eq. (2), 
reduces to
\begin{equation}
R^i_{\alpha_0, j} R^j_{\beta_0} -
R^i_{\beta_0, j} R^j_{\alpha_0} = 
- R^i_{\gamma_0} F^{\gamma_0}_{\alpha_0 \beta_0};
\end{equation}
here, for the sake of simplicity, we assume that the $A^i$ are {\it bosonic} 
fields. Secondly, due to the condition of first--stage reducibility,
\begin{equation}
R^i_{\alpha_0} Z^{\alpha_0}_{\alpha_1} = 0,
\end{equation}
any equation of the form $R^i_{\alpha_0} X^{\alpha_0} = 0$ has the solution 
$X^{\alpha_0} = Z^{\alpha_0}_{\alpha_1} Y^{\alpha_1}$ (for irreducible 
theories Eq. (23) has only the solution $X^{\alpha_0} = 0$). In the case of 
field--dependent structure functions the Jacobi identity takes 
the form
\begin{equation}
R^j_{\delta_0} \bigr(
F^{\delta_0}_{\eta_0 \alpha_0} F^{\eta_0}_{\beta_0 \gamma_0} -
R^i_{\alpha_0} F^{\delta_0}_{\beta_0 \gamma_0, i} +
\hbox{cyclic perm} (\alpha_0, \beta_0, \gamma_0) \bigr) = 0,  
\end{equation}
where the expression in the parenthesis vanishes only for irreducible
theories. It should be noted that the generators $R^i_{\alpha_0}$ and
the zero modes $Z^{\alpha_0}_{\alpha_1}$ are not uniquely defined. By taking 
nonsingular linear combinations of them they can be transformed into the 
so--called standard basis defined in Ref. \cite{3}. But in the following we 
will choose an arbitrary basis without any restriction and proceed 
along the lines of Ref. \cite{8}. 

Let us restrict our considerations to solutions $S_m$ of the {\it classical}
master equations 
\begin{equation*}
\hbox{$\frac{1}{2}$} ( S_m, S_m )^a + V_m^a S_m = 0, 
\qquad
\hbox{$\frac{1}{2}$} \{ S_m, S_m \}_\alpha + V_\alpha S_m = 0,
\end{equation*}
being {\it linear} in the antifields. These equations, because of the 
linearity with respect to the antifields, may be expressed also by 
$\mathbf{s}_m^a S_m = 0$ and $\mathbf{d}_\alpha S_m = 0$, where the symmetry 
operators being denoted by $\mathbf{s}_m^a = \mathbf{s}_m^a \phi^A 
\delta_L/ \delta \phi^A + V_m^a$ and 
$\mathbf{d}_\alpha = \mathbf{d}_\alpha \phi^A 
\delta_L/ \delta \phi^A + V_\alpha$, where $\mathbf{s}_m^a \phi^A =
(-1)^{\epsilon_A} \delta S_m/ \delta \phi_{A a}^*$ and
$\mathbf{d}_\alpha \phi^A = (-1)^{\epsilon_A} (\sigma_\alpha)_B^{~~\!A}
\delta S_m/ \delta \eta_B$, are required to fulfil the 
$osp(1,2)$--superalgebra:
\begin{equation}
[ \mathbf{d}_\alpha, \mathbf{d}_\beta ] = 
\epsilon_{\alpha\beta}^{~~~\!\gamma} \mathbf{d}_\gamma, 
\qquad
[ \mathbf{d}_\alpha, \mathbf{s}_m^a ] = 
\mathbf{s}_m^b (\sigma_\alpha)_b^{~a},
\qquad
\{ \mathbf{s}_m^a, \mathbf{s}_m^b \} = - 
m^2 (\sigma^\alpha)^{ab} \mathbf{d}_\alpha.
\end{equation}
In Ref. \cite{1} it has been shown that such solutions can be written in 
the form 
\begin{equation}
S_m = S_{\rm cl} + 
(\hbox{$\frac{1}{2}$} \epsilon_{ab} \mathbf{s}_m^b \mathbf{s}_m^a + m^2) X,
\end{equation}
where for the first--stage reducible case the $Sp(2)$--scalar $X$ 
has to be choosen as $X = {\bar A}_i A^i + {\bar B}_{\alpha_0} B^{\alpha_0} +
{\bar B}_{\alpha_1 a} B^{\alpha_1 a} + {\bar C}_{\alpha_0 a} C^{\alpha_0 a} +
{\bar C}_{\alpha_1 ab} C^{\alpha_1 ab}$. 
Let us emphasize that $\mathbf{s}_m^a$ and $\mathbf{d}_\alpha$ are {\it not}
related to the first--order differential
operators $\mathbf{Q}_m^a = ( S_m, ~ )^a - i \hbar \bar{\Delta}_m^a$,
$\bar{\Delta}_m^a = \Delta^a + (i/\hbar) V_m^a$ and
$\mathbf{Q}_\alpha = \{ S_m, ~ \}_\alpha - i \hbar \bar{\Delta}_\alpha$,
$\bar{\Delta}_\alpha = \Delta_\alpha + (i/\hbar) V_\alpha$ at the lowest
order approximation of $\hbar$, which was also introduced in Ref. \cite{1}, 
rather they are (nonlinear) realizations of the $osp(1,2)$--superalgebra
in terms of fields and antifields. A realization of the (anti)BRST-- and 
$Sp(2)$--transformations of the antifields already has been given 
(see Appendix B). Thus, we are left with the problem to determine the 
corresponding transformations for the fields $A^i$, 
$B^{\alpha_0}$, $B^{\alpha_1 a}$, $C^{\alpha_0 a}$, $C^{\alpha_1 ab}$.

To begin with let us cast the Jacobi identity (24) into a more practical 
form. Owing to (23) the expression in paranthesis must be proportinal to 
the zero--modes $Z^{\delta_0}_{\alpha_1}$,  
\begin{equation}
F^{\delta_0}_{\eta_0 \alpha_0} F^{\eta_0}_{\beta_0 \gamma_0} -
R^i_{\alpha_0} F^{\delta_0}_{\beta_0 \gamma_0, i} +
\hbox{cyclic perm} (\alpha_0, \beta_0, \gamma_0) =
3 Z^{\delta_0}_{\alpha_1} H^{\alpha_1}_{\alpha_0 \beta_0 \gamma_0},
\end{equation}
where $H^{\alpha_1}_{\alpha_0 \beta_0 \gamma_0}(A)$ are some new structure
functions, being totally antisymmetric 
with respect to the indicies $\alpha_0$, $\beta_0$, $\gamma_0$ and depending, 
in general, on the gauge fields $A^i$. For later use we need an expression 
for the combination $R^j_{\beta_0} Z^{\alpha_0}_{\alpha_1, j}$. Multiplying 
(22) by $Z^{\alpha_0}_{\alpha_1}$ and using the relation 
$R^i_{\alpha_0, j} Z^{\alpha_0}_{\alpha_1} =
- R^i_{\alpha_0} Z^{\alpha_0}_{\alpha_1, j}$, which follows from (23), we get
\begin{equation*}
R^i_{\alpha_0} (
Z^{\alpha_0}_{\alpha_1, j} R^j_{\beta_0} +
F^{\alpha_0}_{\beta_0 \gamma_0} Z^{\gamma_0}_{\alpha_1} ) = 0.
\end{equation*}
Introducing additional new structure functions 
$G^{\gamma_1}_{\beta_0 \alpha_1}(A)$ the solution of the previous 
relation can be written in the form 
\begin{equation}
Z^{\alpha_0}_{\alpha_1, j} R^j_{\beta_0} +
F^{\alpha_0}_{\beta_0 \gamma_0} Z^{\gamma_0}_{\alpha_1} = 
- Z^{\alpha_0}_{\gamma_1} G^{\gamma_1}_{\beta_0 \alpha_1},
\end{equation}
which is a new gauge structure equation for the first--stage reducible
case. Multiplying this equation by $Z^{\beta_0}_{\beta_1}$ and taking into 
account the reducibility condition (23),
\begin{equation}
F^{\alpha_0}_{\beta_0 \gamma_0}
Z^{\gamma_0}_{\alpha_1} Z^{\beta_0}_{\beta_1} =
- Z^{\alpha_0}_{\gamma_1} Z^{\beta_0}_{\beta_1} 
G^{\gamma_1}_{\beta_0 \alpha_1}, 
\end{equation}
for $G^{\gamma_1}_{\beta_0 \alpha_1}$ we obtain the useful
equality
\begin{equation}
Z^{\alpha_0}_{\beta_1} G^{\gamma_1}_{\alpha_0 \alpha_1} = 
- Z^{\alpha_0}_{\alpha_1} G^{\gamma_1}_{\alpha_0 \beta_1}.
\end{equation}
Moreover, using the relation (28), by virtue of (22) and (27) we are able 
to establish two further new gauge structure relations for the first--stage 
reducible case (see Appendix A):
\begin{equation}
\bigr(
G^{\alpha_1}_{\beta_0 \gamma_1} G^{\gamma_1}_{\gamma_0 \beta_1} +
R^i_{\beta_0} G^{\alpha_1}_{\gamma_0 \beta_1, i} +
\hbox{antisym}(\beta_0 \leftrightarrow \gamma_0) \bigr) +
G^{\alpha_1}_{\alpha_0 \beta_1} F^{\alpha_0}_{\beta_0 \gamma_0} +
3 Z^{\alpha_0}_{\beta_1} H^{\alpha_1}_{\alpha_0 \beta_0 \gamma_0} = 0
\end{equation}
and the total antisymmetric expression in 
$(\alpha_0, \beta_0, \gamma_0, \delta_0)$,
\begin{align}
\bigr(& 
H^{\alpha_1}_{\eta_0 \alpha_0 \beta_0} F^{\eta_0}_{\gamma_0 \delta_0} - 
H^{\alpha_1}_{\eta_0 \delta_0 \alpha_0} F^{\eta_0}_{\beta_0 \gamma_0} +
\hbox{cyclic perm} (\alpha_0, \beta_0, \gamma_0) \bigr) 
\nonumber
\\ 
& + \bigr\{ R^i_{\delta_0} H^{\alpha_1}_{\alpha_0 \beta_0 \gamma_0, i} -
G^{\alpha_1}_{\delta_0 \beta_1} H^{\beta_1}_{\alpha_0 \beta_0 \gamma_0} +
\hbox{antisym}\bigr(\delta_0 \leftrightarrow 
(\alpha_0, \beta_0, \gamma_0) \bigr) \bigr\} = 0.
\end{align}
The first one agrees with Eq. (14) in Ref. \cite{5}, but the second one 
differs from Eq. (15) in\break Ref. \cite{5} by terms arising from 
antisymmetrization.  
As a consequence of this difference quartic (anti)ghost terms do not disappear
in the (anti)BRST transformations (see Eq. (39) below). From (31) and (32) 
it follows that the new tensors $H^{\alpha_1}_{\alpha_0 \beta_0 \gamma_0}$ 
and $G^{\alpha_1}_{\alpha_0 \beta_1}$ are not independent of each other.
The gauge commutator relation (22), the Jacobi identity (27) and the new 
gauge structure relations (28), (31) and (32) are the key equations for the 
following considerations.

Let us now derive the (anti)BRST transformations of the fields under 
consideration. Imposing the $osp(1,2)$--superalgebra (25) on the gauge fields 
$A^i$, owing to $\mathbf{d}_\alpha A^i = 0$, this yields 
$\{ \mathbf{s}_m^a, \mathbf{s}_m^b \} A^i = 0$. Then, with
\begin{equation}
\mathbf{s}_m^a A^i = R^i_{\alpha_0} C^{\alpha_0 a},
\end{equation}
and by virtue of (27), we find
\begin{equation*}
R^i_{\alpha_0} (
\mathbf{s}_m^a C^{\alpha_0 b} + \mathbf{s}_m^b C^{\alpha_0 a} +
F^{\alpha_0}_{\beta_0 \gamma_0} C^{\beta_0 a} C^{\gamma_0 b} ) = 0.
\end{equation*} 
The general solution of this equation is
\begin{equation}
\mathbf{s}_m^a C^{\alpha_0 b} = Z^{\alpha_0}_{\alpha_1} C^{\alpha_1 ab} +
\epsilon^{ab} B^{\alpha_0} -
\hbox{$\frac{1}{2}$} F^{\alpha_0}_{\beta_0 \gamma_0} 
C^{\beta_0 a} C^{\gamma_0 b},
\end{equation}
where the (bosonic) ghosts $C^{\alpha_1 ab}$ can be taken to be symmetric,
$C^{\alpha_1 ab} = C^{\alpha_1 ba}$, because its antisymmetric part
enters into the definition of $B^{\alpha_0}$.

Imposing the superalgebra (25) on the (anti)ghosts $C^{\alpha_0 c}$
and taking into account $\mathbf{d}_\alpha C^{\alpha_0 b} =
C^{\alpha_0 c} (\sigma_\alpha)_c^{~b}$ it gives 
$\{ \mathbf{s}_m^a, \mathbf{s}_m^b \} C^{\alpha_0 c} = 
- m^2 (\sigma^\alpha)^{ab} C^{\alpha_0 d} (\sigma_\alpha)_d^{~c}$. The 
right--hand side of this restriction can be rewritten by means of the relations 
$(\sigma_\alpha)_d^{~c} = \epsilon_{de} \epsilon^{fc}
(\sigma_\alpha)^e_{~f}$ and $(\sigma^\alpha)^{ab} (\sigma_\alpha)^e_{~f} =
- (\epsilon^{ae} \delta^b_f + \epsilon^{be} \delta^a_f)$ as
$\{ \mathbf{s}_m^a, \mathbf{s}_m^b \} C^{\alpha_0 c} = - m^2 
(\epsilon^{ac} C^{\alpha_0 b} + \epsilon^{bc} C^{\alpha_0 a})$. Then, 
with (34), by virtue of (27), we obtain
\begin{align*}
\bigr\{& 
Z^{\alpha_0}_{\alpha_1} (
\mathbf{s}_m^a C^{\alpha_1 bc} +  
\hbox{$\frac{1}{2}$} H^{\alpha_1}_{\beta_0 \gamma_0 \delta_0}
C^{\beta_0 a} C^{\gamma_0 b} C^{\delta_0 c} )
\\
& + \epsilon^{bc} \bigr(
\mathbf{s}_m^a B^{\alpha_0} + m^2 C^{\alpha_0 a} -
\hbox{$\frac{1}{2}$} F^{\alpha_0}_{\beta_0 \gamma_0} 
B^{\beta_0} C^{\gamma_0 a}
\\
& - \hbox{$\frac{1}{12}$} \epsilon_{de} (
F^{\alpha_0}_{\eta_0 \beta_0} F^{\eta_0}_{\gamma_0 \delta_0} +
2 R^i_{\beta_0} F^{\alpha_0}_{\gamma_0 \delta_0, i} ) 
C^{\gamma_0 a} C^{\delta_0 d} C^{\beta_0 e} \bigr)
\\
& + \hbox{$\frac{1}{2}$} 
F^{\alpha_0}_{\beta_0 \gamma_0} Z^{\gamma_0}_{\alpha_1} 
C^{\beta_0 c} C^{\alpha_1 ab} + 
( R^i_{\beta_0} Z^{\alpha_0}_{\alpha_1, i} + \hbox{$\frac{1}{2}$} 
F^{\alpha_0}_{\beta_0 \gamma_0} Z^{\gamma_0}_{\alpha_1} ) 
C^{\beta_0 a} C^{\alpha_1 bc} \bigr\} +
\hbox{\rm sym}(a \leftrightarrow b) = 0.
\end{align*}
Replacing $R^i_{\beta_0} Z^{\alpha_0}_{\alpha_1, i}$ according to (28) and 
using the relation
\begin{align*}
\hbox{$\frac{1}{2}$} 
F^{\alpha_0}_{\beta_0 \gamma_0} \bigr\{ 
Z^{\gamma_0}_{\alpha_1} (
C^{\beta_0 c} C^{\alpha_1 ab} - C^{\beta_0 a} C^{\alpha_1 bc} ) -
\epsilon^{bc} \epsilon_{de} Z^{\beta_0}_{\alpha_1} 
C^{\alpha_1 ad} C^{\gamma_0 e} \bigr\} +
\hbox{\rm sym}(a \leftrightarrow b) = 0; 
\end{align*}
this leads to
\begin{align*}
\bigr\{& 
Z^{\alpha_0}_{\alpha_1} (
\mathbf{s}_m^a C^{\alpha_1 bc} -
G^{\alpha_1}_{\beta_0 \beta_1} C^{\beta_0 a} C^{\beta_1 bc} +
\hbox{$\frac{1}{2}$} H^{\alpha_1}_{\beta_0 \gamma_0 \delta_0}
C^{\beta_0 a} C^{\gamma_0 b} C^{\delta_0 c} ) 
\\
& + \epsilon^{bc} \bigr[
\mathbf{s}_m^a B^{\alpha_0} + m^2 C^{\alpha_0 a} -
\hbox{$\frac{1}{2}$} F^{\alpha_0}_{\beta_0 \gamma_0} (
B^{\beta_0} C^{\gamma_0 a} -
\epsilon_{de} Z^{\beta_0}_{\alpha_1} C^{\alpha_1 ad} C^{\gamma_0 e} ) 
\\
& - \hbox{$\frac{1}{12}$} \epsilon_{de} (
F^{\alpha_0}_{\eta_0 \beta_0} F^{\eta_0}_{\gamma_0 \delta_0} +
2 R^i_{\beta_0} F^{\alpha_0}_{\gamma_0 \delta_0, i} )
C^{\gamma_0 a} C^{\delta_0 d} C^{\beta_0 e} \bigr] \bigr\} +
\hbox{\rm sym}(a \leftrightarrow b) = 0.
\end{align*}
Here, $\mathbf{s}_m^a B^{\alpha_0}$ can give a local contribution to
$\mathbf{s}_m^a C^{\alpha_1 bc}$ if and only if it is proportional to 
$Z^{\alpha_0}_{\alpha_1}$. Therefore, if we introduce with 
$\mathbf{s}_m^a B^{\alpha_0}$ the new (fermionic) auxiliary field 
$B^{\alpha_1 a}$ according to
\begin{align}
\mathbf{s}_m^a B^{\alpha_0} &= 
Z^{\alpha_0}_{\alpha_1} B^{\alpha_1 a} +
\hbox{$\frac{1}{2}$} F^{\alpha_0}_{\beta_0 \gamma_0} (
B^{\beta_0} C^{\gamma_0 a} -
\epsilon_{cd} Z^{\beta_0}_{\alpha_1} C^{\alpha_1 ac} C^{\gamma_0 d})
\nonumber\\
& \quad
+ \hbox{$\frac{1}{12}$} \epsilon_{cd} (
F^{\alpha_0}_{\eta_0 \beta_0} F^{\eta_0}_{\gamma_0 \delta_0} +
2  R^i_{\beta_0} F^{\alpha_0}_{\gamma_0 \delta_0, i} )
C^{\gamma_0 a} C^{\delta_0 c} C^{\beta_0 d} - 
m^2 C^{\alpha_0 a},  
\end{align}
for $\mathbf{s}_m^a C^{\alpha_1 bc}$ we will get the equation
\begin{equation*}
Z^{\alpha_0}_{\alpha_1} \bigr\{
\mathbf{s}_m^a C^{\alpha_1 bc} + \epsilon^{bc} B^{\alpha_1 a} - 
G^{\alpha_1}_{\beta_0 \beta_1} C^{\beta_0 a} C^{\beta_1 bc} +
\hbox{$\frac{1}{2}$} H^{\alpha_1}_{\beta_0 \gamma_0 \delta_0}
C^{\beta_0 a} C^{\gamma_0 b} C^{\gamma_0 c} \bigr\} +
\hbox{\rm sym}(a \leftrightarrow b) = 0.
\end{equation*}
Because the ghosts $C^{\alpha_1 bc}$ are symmetric with respect to $b$ and $c$
the general solution of this equation is of the form
\begin{equation}
\mathbf{s}_m^a C^{\alpha_1 bc} =
- \epsilon^{ac} B^{\alpha_1 b} - \epsilon^{ab} B^{\alpha_1 c} + 
G^{\alpha_1}_{\alpha_0 \beta_1} C^{\alpha_0 a} C^{\beta_1 bc} -
\hbox{$\frac{1}{2}$} H^{\alpha_1}_{\alpha_0 \beta_0 \gamma_0}
C^{\alpha_0 a} C^{\beta_0 b} C^{\gamma_0 c}. 
\end{equation}
The expression for $\mathbf{s}_m^a B^{\alpha_1 b}$ can be found by applying 
the superalgebra (25) on $B^{\alpha_0}$. Due to 
$\mathbf{d}_\alpha B^{\alpha_0} = 0$, this leads to the requirement
$\{ \mathbf{s}_m^a, \mathbf{s}_m^b \} B^{\alpha_0} = 0$. After a somewhat
involved algebraic calculation this gives
\begin{align*}
\bigr\{&
Z^{\lambda_0}_{\alpha_1} \bigr[
\mathbf{s}_m^a B^{\alpha_1 b} - m^2 C^{\alpha_1 ab} -
G^{\alpha_1}_{\alpha_0 \beta_1} C^{\alpha_0 a} B^{\beta_1 b} 
\\
& + H^{\alpha_1}_{\alpha_0 \beta_0 \gamma_0} (
\hbox{$\frac{1}{2}$} B^{\alpha_0} C^{\beta_0 a} C^{\gamma_0 b} - 
\epsilon_{cd} C^{\beta_0 a} 
Z^{\gamma_0}_{\beta_1} C^{\beta_1 bc} C^{\alpha_0 d} )
\\
& - \hbox{$\frac{1}{8}$} \epsilon_{cd} (
G^{\alpha_1}_{\delta_0 \beta_1} H^{\beta_1}_{\alpha_0 \beta_0 \gamma_0} -
R^i_{\delta_0} H^{\alpha_1}_{\alpha_0 \beta_0 \gamma_0, i} ) 
C^{\gamma_0 a} C^{\beta_0 b} C^{\alpha_0 c} C^{\delta_0 d} 
\\
& + \hbox{$\frac{1}{8}$} \epsilon_{cd}
H^{\alpha_1}_{\eta_0 \alpha_0 \beta_0} F^{\eta_0}_{\gamma_0 \delta_0} 
C^{\gamma_0 a} C^{\beta_0 b} C^{\alpha_0 c} C^{\delta_0 d} \bigr] + 
\hbox{$\frac{1}{2}$} \epsilon_{cd} F^{\lambda_0}_{\alpha_0 \beta_0} 
Z^{\alpha_0}_{\alpha_1} C^{\alpha_1 ac} 
Z^{\beta_0}_{\beta_1} C^{\beta_1 bd} \bigr\} + 
\hbox{\rm sym}(a \leftrightarrow b) = 0,
\end{align*}
and further, by virtue of (29),
\begin{align}
Z^{\lambda_0}_{\alpha_1} \bigr\{& 
\mathbf{s}_m^a B^{\alpha_1 b} - m^2 C^{\alpha_1 ab} -
G^{\alpha_1}_{\alpha_0 \beta_1} C^{\alpha_0 a} B^{\beta_1 b} 
\nonumber\\
& + H^{\alpha_1}_{\alpha_0 \beta_0 \gamma_0} (
\hbox{$\frac{1}{2}$} B^{\alpha_0} C^{\beta_0 a} C^{\gamma_0 b} - 
\epsilon_{cd} C^{\beta_0 a} 
Z^{\gamma_0}_{\beta_1} C^{\beta_1 bc} C^{\alpha_0 d} )
\\
& - \hbox{$\frac{1}{8}$} \epsilon_{cd} (
G^{\alpha_1}_{\delta_0 \beta_1} H^{\beta_1}_{\alpha_0 \beta_0 \gamma_0} -
R^i_{\delta_0} H^{\alpha_1}_{\alpha_0 \beta_0 \gamma_0, i} ) 
C^{\gamma_0 a} C^{\beta_0 b} C^{\alpha_0 c} C^{\delta_0 d} 
\nonumber\\
& + \hbox{$\frac{1}{8}$} \epsilon_{cd}
H^{\alpha_1}_{\eta_0 \alpha_0 \beta_0} F^{\eta_0}_{\gamma_0 \delta_0} 
C^{\gamma_0 a} C^{\beta_0 b} C^{\alpha_0 c} C^{\delta_0 d} + 
\hbox{$\frac{1}{2}$} \epsilon_{cd} G^{\alpha_1}_{\alpha_0 \beta_1} 
Z^{\alpha_0}_{\gamma_1} C^{\gamma_1 ac} C^{\beta_1 bd} \bigr\} + 
\hbox{\rm sym}(a \leftrightarrow b) = 0.
\nonumber
\end{align}
In deriving the cubic (anti)ghost terms in this equation the following 
equality was used:
\begin{equation*}
\epsilon_{ab} \delta^d_c + 
\epsilon_{bc} \delta^d_a + \epsilon_{ca} \delta^d_b = 0.
\end{equation*}
Let us point out that in (37) the quartic (anti)ghost terms cannot be 
droped due to our modification of relation (32).

Another equation for $\mathbf{s}_m^a B^{\alpha_1 b}$ can be obtained by 
applying the superalgebra (25) on $C^{\alpha_1 cd}$. Due to 
$\mathbf{d}_\alpha C^{\alpha_1 cd} = C^{\alpha_1 ed} (\sigma_\alpha)_e^{~c} +
C^{\alpha_1 ce} (\sigma_\alpha)_e^{~d}$, this leads to 
$\{ \mathbf{s}_m^a, \mathbf{s}_m^b \} C^{\alpha_1 cd} = - m^2 
(\sigma^\alpha)^{ab} \bigr( C^{\alpha_1 ed} (\sigma_\alpha)_e^{~c} +
C^{\alpha_1 ce} (\sigma_\alpha)_e^{~d} \bigr)$. The rigth--hand side of this
equation can be re--\break written as 
$\{ \mathbf{s}_m^a, \mathbf{s}_m^b \} C^{\alpha_1 cd} = - m^2 
( \epsilon^{ac} C^{\alpha_1 bd} + \epsilon^{ad} C^{\alpha_1 bc} +
\epsilon^{bc} C^{\alpha_1 ad} + \epsilon^{bd} C^{\alpha_1 ac} )$. Then, with
(34), taking into account the relation (31), we obtain  
\begin{align}
\bigr\{& 
\epsilon^{bc} (
\mathbf{s}_m^a B^{\alpha_1 d} - m^2 C^{\alpha_1 ad} -
G^{\alpha_1}_{\alpha_0 \beta_1} C^{\alpha_0 a} B^{\beta_1 d} +
\hbox{$\frac{1}{2}$} H^{\alpha_1}_{\alpha_0 \beta_0 \gamma_0}
B^{\alpha_0} C^{\beta_0 a} C^{\gamma_0 d} ) +
\hbox{\rm sym}(c \leftrightarrow d) 
\nonumber\\
& - G^{\alpha_1}_{\alpha_0 \beta_1} 
Z^{\alpha_0}_{\gamma_1} C^{\gamma_1 ab} C^{\beta_1 cd} -
\hbox{$\frac{3}{2}$} H^{\alpha_1}_{\alpha_0 \beta_0 \gamma_0} 
Z^{\alpha_0}_{\beta_1} C^{\beta_1 cd} C^{\beta_0 a} C^{\gamma_0 b}   
\nonumber\\
& + \hbox{$\frac{1}{2}$} 
H^{\alpha_1}_{\alpha_0 \beta_0 \gamma_0} Z^{\alpha_0}_{\beta_1} (
C^{\beta_1 ab} C^{\beta_0 c} C^{\gamma_0 d} +
C^{\beta_1 ac} C^{\beta_0 d} C^{\gamma_0 b} +
C^{\beta_1 ad} C^{\beta_0 b} C^{\gamma_0 c} )
\\
& - \hbox{$\frac{1}{2}$} (
G^{\alpha_1}_{\delta_0 \beta_1} H^{\beta_1}_{\alpha_0 \beta_0 \gamma_0} -
R^i_{\delta_0} H^{\alpha_1}_{\alpha_0 \beta_0 \gamma_0, i} ) 
C^{\delta_0 a} C^{\alpha_0 b} C^{\beta_0 c} C^{\gamma_0 d}
\nonumber\\
& + \hbox{$\frac{1}{4}$} 
H^{\alpha_1}_{\eta_0 \alpha_0 \beta_0} F^{\eta_0}_{\gamma_0 \delta_0} 
C^{\delta_0 a} (
C^{\alpha_0 b} C^{\beta_0 c} C^{\gamma_0 d} +
C^{\beta_0 b} C^{\gamma_0 c} C^{\alpha_0 d} +
C^{\gamma_0 b} C^{\alpha_0 c} C^{\beta_0 d} ) \bigr\} +
\hbox{\rm sym}(a \leftrightarrow b) = 0, 
\nonumber
\end{align}
where again, for the same reason as before, quartic (anti)ghost cannot be 
droped. 

It is not very difficult to see, by virtue of (32), that the general solution 
of (37) and (38) reads 
\begin{align}
\mathbf{s}_m^a B^{\alpha_1 b} &= G^{\alpha_1}_{\alpha_0 \beta_1} (
C^{\alpha_0 a} B^{\beta_1 b} - 
\hbox{$\frac{1}{2}$} \epsilon_{cd} 
Z^{\alpha_0}_{\gamma_1} C^{\gamma_1 ac} C^{\beta_1 bd} ) - 
\hbox{$\frac{1}{2}$} H^{\alpha_1}_{\alpha_0 \beta_0 \gamma_0} 
B^{\alpha_0} C^{\beta_0 a} C^{\gamma_0 b}
\nonumber\\
& \quad ~ + \hbox{$\frac{1}{4}$} \epsilon_{cd}
H^{\alpha_1}_{\alpha_0 \beta_0 \gamma_0} Z^{\alpha_0}_{\beta_1} (
3 C^{\beta_0 a} C^{\beta_1 bc} C^{\gamma_0 d} +
C^{\beta_0 b} C^{\beta_1 ac} C^{\gamma_0 d} )
\\
& \quad~ + \hbox{$\frac{1}{8}$} \epsilon_{cd} (
G^{\alpha_1}_{\delta_0 \beta_1} H^{\beta_1}_{\alpha_0 \beta_0 \gamma_0} -
R^i_{\delta_0} H^{\alpha_1}_{\alpha_0 \beta_0 \gamma_0, i} )
C^{\gamma_0 a} C^{\beta_0 b} C^{\alpha_0 c} C^{\delta_0 d}
\nonumber\\
& \quad~ - \hbox{$\frac{1}{16}$} \epsilon_{cd} 
H^{\alpha_1}_{\eta_0 \alpha_0 \beta_0} F^{\eta_0}_{\gamma_0 \delta_0} 
( C^{\gamma_0 a} C^{\beta_0 b} + C^{\gamma_0 b} C^{\beta_0 a} ) 
C^{\alpha_0 c} C^{\delta_0 d} + m^2 C^{\alpha_1 ab}. 
\nonumber
\end{align}
Because in (39) no new auxiliary fields had to be introduced one would
expect that the condition 
$\{ \mathbf{s}_m^a, \mathbf{s}_m^b \} B^{\alpha_1 c} =
- m^2 (\sigma^\alpha)^{ab} B^{\alpha_1 d} (\sigma_\alpha)_d^{~c}$ should be
fulfilled identically as a consequence of the previous formalae. 
Corresponding direct calculations require the same tedious algebraic work but 
it can be proved that this relation is indeed satisfied. 

The relations (33)--(36) and (39) specify the transformations of the 
$osp(1,2)$--symmetry for first--stage reducible massive theories with
closed gauge algebra. By using the method of Ref. \cite{3} it can be shown
that the solution $S_m$, Eq. (26), is the most general one of the classical 
master equations with vanishing new ghost number, i. e., ${\rm ngh}(S_m) = 0$.  

Finally, let us determine the action $S_{m, {\rm eff}}$ in the vacuum
functional (19) for the class of {\it minimal} gauges $F$ depending only on 
the fields $A^i$, the (anti)ghosts $C^{\alpha_0 a}$, $C^{\alpha_1 ab}$ and the 
auxiliary fields $B^{\alpha_0}$, $B^{\alpha_1 a}$. Inserting into (20) for 
$S_m$ the action 
\begin{align*}
S_m &= S_{\rm cl} + A_{ia}^* (\mathbf{s}_m^a A^i) +
\bar{A}_i ( \hbox{$\frac{1}{2}$} \epsilon_{ab} 
\mathbf{s}_m^b \mathbf{s}_m^a A^i ) + 
B_{\alpha_0 a}^* (\mathbf{s}_m^a B^{\alpha_0}) +
\bar{B}_{\alpha_0} ( \hbox{$\frac{1}{2}$} \epsilon_{ab} 
\mathbf{s}_m^b \mathbf{s}_m^a B^{\alpha_0} )  
\\
& \quad~ + F_{\alpha_0 c} C^{\alpha_0 c} - 
C_{\alpha_0 ac}^* (\mathbf{s}_m^a C^{\alpha_0 c}) +
\bar{C}_{\alpha_0 c} ( \hbox{$\frac{1}{2}$} \epsilon_{ab} 
\mathbf{s}_m^b \mathbf{s}_m^a C^{\alpha_0 c} ) -
\hbox{$\frac{1}{2}$} m^2 \bar{C}_{\alpha_0 c} C^{\alpha_0 c} 
\\
& \quad~ + E_{\alpha_1 a} B^{\alpha_1 a} - 
B_{\alpha_1 ab}^* (\mathbf{s}_m^a B^{\alpha_1 b}) +
\bar{B}_{\alpha_1 c} ( \hbox{$\frac{1}{2}$} \epsilon_{ab} 
\mathbf{s}_m^b \mathbf{s}_m^a B^{\alpha_1 c} ) -
\hbox{$\frac{1}{2}$} m^2 \bar{B}_{\alpha_1 c} B^{\alpha_1 c}
\\
& \quad~ + F_{\alpha_1 cd} C^{\alpha_1 cd} + 
C_{\alpha_1 acd}^* (\mathbf{s}_m^a C^{\alpha_1 cd}) +
\bar{C}_{\alpha_1 cd} ( \hbox{$\frac{1}{2}$} \epsilon_{ab} 
\mathbf{s}_m^b \mathbf{s}_m^a C^{\alpha_1 cd} ) -
m^2 \bar{C}_{\alpha_1 cd} C^{\alpha_1 cd}
\end{align*}
and performing the integration over antifields and auxiliary fields we get
the following expression for $S_{m, {\rm eff}}$
(at the lowest order of $\hbar$):
\begin{equation*}
Z_m(0) = \int d A^i\, d B^{\alpha_0}\, d C^{\alpha_0 a}\,
d B^{\alpha_1 a}\, d C^{\alpha_1 ab}\,
{\exp}\{ (i/ \hbar) S_{m, {\rm eff}} \},
\qquad
S_{m, {\rm eff}} = S_{\rm cl} + W_F,
\end{equation*}
where $W_F$ is given by
\begin{align*}
W_F &= \hbox{$\frac{1}{2}$} \epsilon_{ab} (
\frac{\delta F}{\delta A^i}
\mathbf{s}_m^b \mathbf{s}_m^a A^i -
\mathbf{s}_m^a A^i
\frac{\delta^2 F}{\delta A^i \delta A^j} 
\mathbf{s}_m^b A^j ) 
\\
& \quad~ + \hbox{$\frac{1}{2}$} \epsilon_{ab} (
\frac{\delta F}{\delta C^{\alpha_0 c}}
\mathbf{s}_m^b \mathbf{s}_m^a C^{\alpha_0 c} -
\mathbf{s}_m^a C^{\alpha_0 c}  
\frac{\delta^2 F}{\delta C^{\alpha_0 c} \delta C^{\beta_0 d}} 
\mathbf{s}_m^b C^{\beta_0 d} )  
\\
& \quad~ + \hbox{$\frac{1}{2}$} \epsilon_{ab} (
\frac{\delta F}{\delta B^{\alpha_1 c}}
\mathbf{s}_m^b \mathbf{s}_m^a B^{\alpha_1 c} -
\mathbf{s}_m^a B^{\alpha_1 c}
\frac{\delta^2 F}{\delta B^{\alpha_1 c} \delta B^{\beta_1 d}} 
\mathbf{s}_m^b B^{\beta_1 d} ) 
\\
& \quad~ + \hbox{$\frac{1}{2}$} \epsilon_{ab} (
\frac{\delta F}{\delta C^{\alpha_1 cd}}
\mathbf{s}_m^b \mathbf{s}_m^a C^{\alpha_1 cd} -
\mathbf{s}_m^a C^{\alpha_1 cd}
\frac{\delta^2 F}{\delta C^{\alpha_1 cd} \delta C^{\beta_1 ef}} 
\mathbf{s}_m^b C^{\beta_1 ef} )  
\\
& \quad~ - \hbox{$\frac{1}{2}$} \epsilon_{ab} 
\mathbf{s}_m^a C^{\alpha_0 c}
\frac{\delta^2 F}{\delta C^{\alpha_0 c} \delta B^{\alpha_1 d}} 
\mathbf{s}_m^b B^{\alpha_1 d} + m^2 F. 
\end{align*}
This gauge--fixing term can be rewritten as
\begin{equation*}
W_F = ( \hbox{$\frac{1}{2}$} \epsilon_{ab}
\mathbf{s}_m^b \mathbf{s}_m^a + m^2 ) F,
\end{equation*}
showing that the action $S_{m, {\rm eff}}$ is in fact $osp(1,2)$--invariant 
and that the method of gauge fixing suggested in Section II will actually 
remove the degeneracy of the classical action.
\bigskip\medskip
\begin{flushleft}
{\large{\bf VI. EXAMPLES}} 
\end{flushleft}
\bigskip
As a first example let us give the $osp(1,2)$--symmetric generalization of
the Chapline--Manton model \cite{7} which describes the unified $N = 1$ 
supersymmetric Yang--Mills theory and $N = 1$ supergravity in ten dimensions.
A striking feature of this model is that the Yang--Mills part of the 
classical action (the dots $\cdots$ indicate the supergravity part and 
additional terms of the super--Yang--Mills part) 
\begin{gather*}
S_{\rm CM} = - \hbox{$\frac{3}{4}$} (
\partial_{[\rho} A_{\mu\nu]} - X_{\rho\mu\nu} ) (
\partial^{[\rho} A^{\mu\nu]} - X^{\rho\mu\nu} ) + \cdots,
\\
X_{\rho\mu\nu} \equiv A_{[\rho}^\alpha G_{\mu\nu]}^\alpha -
\hbox{$\frac{1}{3}$} F^{\alpha\beta\gamma}
A_{[\rho}^\alpha A_\mu^\beta A_{\nu]}^\gamma,
\qquad
G_{\mu\nu}^\alpha \equiv \partial_{[\mu} A_{\nu]}^\alpha +
F^{\alpha\beta\gamma} A_\mu^\beta A_\nu^\gamma,
\end{gather*}
exhibits a new type of (mixed) gauge invariance; here $X_{\rho\mu\nu}$ is the 
Chern--Simons 3--form, $G_{\mu\nu}^\alpha$ the ordinary Yang--Mills field 
strength and $F^{\alpha\beta\gamma}$ are the totally antisymmetric
structure constants. The non--abelian gauge transformation of the Yang--Mills 
potential $A_\mu^{\alpha}$ is accompanied by an abelian gauge transformation 
of the skew symmetric supergravity potential $A_{\mu\nu}$:
\begin{equation*}
\delta A_\mu^\alpha = D_\mu^{\alpha \beta} \theta^\beta(x),
\qquad
D_\mu^{\alpha \beta} \equiv \delta^{\alpha \beta} \partial_\mu -
F^{\alpha \beta \gamma} A_\mu^{\gamma},
\qquad
\delta A_{\mu\nu} = \partial_{[\mu} \theta_{\nu]}(x) +
\theta^\alpha(x) \partial_{[\mu} A_{\nu]}^\alpha,
\end{equation*}
where we have droped the gauge and supergravity coupling constant; in addition 
$A_\mu^{\alpha}$ and $A_{\mu\nu}$ undergo supersymmetric transformations. 
This theory is a first--stage reducible one with closed gauge algebra. Its 
complete spectrum of (anti)ghosts and auxiliary fields, $C^{\alpha a}$,
$B^\alpha$ and $C_\mu^a$, $B_\mu$, $C^{ab}$, $B^a$, has been constructed 
in Ref. \cite{12}. In order to obtain the $osp(1,2)$--symmetric generalization 
of the corresponding {\it massive} theory the gauge--fixed action will be 
written as  
\begin{equation*}
S_{m, {\rm eff}} = S_{\rm CM} + (
\hbox{$\frac{1}{2}$} \epsilon_{ab} \mathbf{s}_m^b \mathbf{s}_m^a + m^2 ) F
\end{equation*}
with gauge fixing functional
\begin{equation*}
F = \hbox{$\frac{1}{2}$} (
A_\mu^\alpha A^{\mu \alpha} + \xi \epsilon_{ab} C^{\alpha a} C^{\alpha b} ) + 
\hbox{$\frac{1}{2}$} \tau \bigr(
A_{\mu \nu} A^{\mu \nu} + \rho (
\epsilon_{cd} C_\mu^c C^{\mu d} +
\hbox{$\frac{1}{2}$} \sigma \epsilon_{cd} \epsilon_{ef} C^{ce} C^{df} ) 
\bigr) + \cdots,
\end{equation*}
$\xi$, $\rho$, $\sigma$ and $\tau$ being the gauge parameters, where the
dots $\cdots$ stand for all usually necessary terms for fixing the
supergravity gauge \cite{13}. For the fields $A_\mu^\alpha$ the
extended BRST transformations has been given in Ref. \cite{1}: 
\begin{align*}
\mathbf{s}_m^a A_\mu^\alpha &= 
D_\mu^{\alpha \beta} C^{\beta a},
\\
\mathbf{s}_m^a C^{\alpha b} &= 
\epsilon^{ab} B^\alpha - \hbox{$\frac{1}{2}$} F^{\alpha \beta \gamma} 
C^{\beta a} C^{\gamma b},
\\
\mathbf{s}_m^a B^\alpha &= - m^2 C^{\alpha a} +
\hbox{$\frac{1}{2}$} F^{\alpha \beta \gamma} B^\beta C^{\gamma a} +
\hbox{$\frac{1}{12}$} \epsilon_{cd} 
F^{\alpha \eta \beta} F^{\eta \gamma \delta} 
C^{\gamma a} C^{\delta c} C^{\beta d}
\\
\intertext{and for the fields $A_{\mu\nu}$ the procedure outlined in 
(22)--(39) yields (in accordance with \cite{5} up to $m$--dependent terms)}
\mathbf{s}_m^a A_{\mu\nu} &= 
\partial_{[\mu} C_{\nu]}^a +
C^{\alpha a} \partial_{[\mu} A_{\nu]}^\alpha,
\\
\mathbf{s}_m^a C_\mu^b &=
\partial_\mu C^{ab} + \epsilon^{ab} B_\mu -
\hbox{$\frac{1}{2}$} F^{\alpha \beta \gamma} 
A_\mu^\alpha C^{\beta a} C^{\gamma b},
\\
\mathbf{s}_m^a B_\mu &=
- m^2 C_\mu^a + \partial_\mu B^a + 
\hbox{$\frac{1}{6}$} \epsilon_{cd} F^{\alpha \beta \gamma} 
C^{\alpha a} C^{\beta c} D_\mu^{\gamma \delta} C^{\delta d}
\\
& \quad~
+ \hbox{$\frac{1}{2}$} F^{\alpha \beta \gamma} 
A_\mu^\alpha B^\beta C^{\gamma a} + 
\hbox{$\frac{1}{12}$} \epsilon_{cd} 
F^{\alpha \eta \beta} F^{\eta \gamma \delta} 
A_\mu^\alpha C^{\gamma a} C^{\delta c} C^{\beta d}, 
\\
\mathbf{s}_m^a C^{bc} &=
- \epsilon^{ab} B^c - \epsilon^{ac} B^b +
\hbox{$\frac{1}{6}$} F^{\alpha \beta \gamma} 
C^{\alpha a} C^{\beta b} C^{\gamma c},
\\
\mathbf{s}_m^a B^b &=
m^2 C^{ab} + \hbox{$\frac{1}{6}$} F^{\alpha \beta \gamma} 
B^\alpha C^{\beta a} C^{\gamma b}.
\end{align*}
The improvement of the results in \cite{5} through Eq. (39) does not matter
here since the corresponding symmetry is abelian.

For the corresponding gauge--fixing terms one gets
\begin{align*}
\hbox{$\frac{1}{4}$} \epsilon_{ab} 
\mathbf{s}_m^b \mathbf{s}_m^a (A_\mu^\alpha A^{\mu \alpha}) &= 
A_\mu^\alpha \partial^\mu B^\alpha +
\hbox{$\frac{1}{2}$} \epsilon_{ab} (\partial^\mu C^{\alpha b})
D_\mu^{\alpha \beta} C^{\beta a},
\\
\hbox{$\frac{1}{4}$} \epsilon_{ab} 
\mathbf{s}_m^b \mathbf{s}_m^a (\epsilon_{cd} C^{\alpha c} C^{\alpha d}) &= 
\hbox{$\frac{1}{2}$} m^2 \epsilon_{cd} C^{\alpha c} C^{\alpha d} +
B^\alpha B^\alpha -
\hbox{$\frac{1}{24}$} \epsilon_{ab} \epsilon_{cd}
F^{\eta \alpha \beta} F^{\eta \gamma \delta}
C^{\alpha a} C^{\beta c} C^{\gamma b} C^{\delta d}
\\
\intertext{and}
\hbox{$\frac{1}{4}$} \epsilon_{ab} 
\mathbf{s}_m^b \mathbf{s}_m^a (A_{\mu\nu} A^{\mu\nu}) &= 
A_{\mu\nu} \partial^{[\mu} B^{\nu]} +
\hbox{$\frac{1}{2}$} \epsilon_{ab} 
( \partial_{[\mu} C_{\nu]}^b + 
C^{\alpha b} \partial_{[\mu} A_{\nu]}^\alpha ) 
( \partial^{[\mu} C^{\nu] a} + 
C^{\alpha a} \partial^{[\mu} A^{\nu] \alpha} ), 
\\
\hbox{$\frac{1}{4}$} \epsilon_{ab}  
\mathbf{s}_m^b \mathbf{s}_m^a (\epsilon_{cd} C_\mu^c C^{\mu d}) &=
\hbox{$\frac{1}{2}$} m^2 \epsilon_{cd} C_\mu^c C^{\mu d} +
B_\mu B^\mu - 2 \epsilon_{cd} (\partial_\mu B^c) C^{\mu d} +
\epsilon_{cd} F^{\alpha \beta \gamma} 
A_\mu^\alpha B^\beta C^{\gamma c} C^{\mu d}
\\
& \quad
+ \hbox{$\frac{1}{2}$} \epsilon_{ab} \epsilon_{cd}
( \partial_\mu C^{ac} - \hbox{$\frac{1}{2}$} F^{\eta \alpha \beta} 
A_\mu^\eta C^{\alpha a} C^{\beta c} )
( \partial^\mu C^{bd} - \hbox{$\frac{1}{2}$} F^{\eta \gamma \delta} 
A^{\mu \eta} C^{\gamma b} C^{\delta d} )
\\
& \quad
+ \hbox{$\frac{1}{3}$} \epsilon_{cd} ( 
F^{\alpha \beta \gamma} 
C^{\alpha a} C^{\beta c} D_\mu^{\gamma \delta} C^{\delta b} +
\hbox{$\frac{1}{2}$} F^{\alpha \eta \beta}
F^{\eta \gamma \delta} A_\mu^\alpha 
C^{\gamma a} C^{\delta c} C^{\beta b} ) C^{\mu d}, 
\\
\hbox{$\frac{1}{8}$} \epsilon_{ab}  
\mathbf{s}_m^b \mathbf{s}_m^a (\epsilon_{cd} \epsilon_{ef} C^{ce} C^{df}) &=
\hbox{$\frac{1}{2}$} m^2 \epsilon_{cd} \epsilon_{ef} C^{ce} C^{df} - 
\hbox{$\frac{3}{2}$} \epsilon_{ab} B^a B^b +
\hbox{$\frac{1}{4}$} \epsilon_{cd} \epsilon_{ef} 
F^{\alpha \beta \gamma} B^\alpha C^{\beta c} C^{\gamma e} C^{df}
\\
& \quad - \hbox{$\frac{1}{144}$} \epsilon_{ab} \epsilon_{cd} \epsilon_{ef} 
( F^{\eta \beta \gamma} C^{\eta a} C^{\beta c} C^{\gamma e} )
( F^{\eta \gamma \delta} C^{\eta b} C^{\gamma d} C^{\delta f} ).
\end{align*}
The elimination of $B^\alpha$, $B_\mu$ and $B^a$ can be performed by
gaussian integration; it provides the gauge--fixing terms 
$\hbox{$\frac{1}{2}$} \xi^{-1} ( \partial^\mu A_\mu^\alpha +
\tau \rho \epsilon_{cd} F^{\alpha\beta\gamma} A_\mu^\beta 
C^{\gamma c} C^{\mu d} )^2$ and $\rho^{-1} ( \partial^\mu A_{\mu\nu} )^2$ 
as well as higher--order (anti)ghost interaction terms. Thus, the degeneracy 
of the classical action is indeed removed. If compared with Ref. \cite{12}, 
here the complete spectrum of (anti)ghosts $C^{\alpha a}$, $C_\mu^a$, $C^{ab}$ 
and auxiliary fields $B^\alpha$, $B_\mu$, $B^a$ is produced in a direct manner 
and, due to the $Sp(2)$--invariance, also a much more compact notation is 
obtained which simplifies all the formulas. 

As a second example let us give the $osp(1,2)$--symmetric generalization 
of {\it topological} Yang--Mills theory \cite{8} in four dimensions (the 
interest in such a theory is its connection to Donaldson theory \cite{14}). 
The classical action is proportional to the Pontryagin index
\begin{equation*}
S_{\rm TYM} = \hbox{$\frac{1}{4}$} G_{\mu\nu}^\alpha \tilde{G}^{\mu\nu\alpha},
\qquad
G_{\mu\nu}^\alpha \equiv \partial_{[\mu} A_{\nu]}^\alpha +
F^{\alpha\beta\gamma} A_\mu^\beta A_\nu^\gamma,
\qquad
\tilde{G}_{\mu\nu}^\alpha \equiv \hbox{$\frac{1}{2}$} 
\epsilon_{\mu\nu\rho\sigma} G^{\rho\sigma\alpha},
\end{equation*}
where $\tilde{G}_{\mu\nu}^\alpha$ is the dual field stength. Since the
Pontryagin index is a group invariant the action is invariant under two 
types of gauge transformations
\begin{equation*}
\delta A_\mu^\alpha = D_\mu^{\alpha\beta} \theta^\beta(x) + 
\theta_\mu^\alpha(x),
\qquad
D_\mu^{\alpha\beta} \equiv \delta^{\alpha\beta} \partial_\mu -
F^{\alpha\beta\gamma} A_\mu^\gamma,
\end{equation*}
which form a closed algebra: the commutator of two gauge transformations 
with parameters ($\rho^\alpha, \rho_\mu^\alpha$) and
($\sigma^\beta, \sigma_\mu^\beta$) corresponds to a gauge transformation
with parameters ($F^{\gamma\alpha\beta} \rho^\alpha \sigma^\beta,
F^{\gamma\alpha\beta} (\rho^\alpha \sigma_\mu^\beta +
\rho_\mu^\alpha \sigma^\beta)$). This theory is a first--stage reducible one 
since obviously both gauge transformations are not independent. Its complete 
spectrum of (anti)ghosts and auxiliary fields, $C^{\alpha a}$, $B^\alpha$ and 
$C_\mu^{\alpha a}$, $B_\mu^\alpha$, $C^{\alpha ab}$, $B^{\alpha a}$, has been 
constructed in Ref. \cite{15}. In order to obtain the $osp(1,2)$--symmetric 
generalization of the corresponding {\it massive} theory the gauge--fixed 
action will be cast into the form  
\begin{equation*}
S_{m, {\rm eff}} = S_{\rm TYM} + (
\hbox{$\frac{1}{2}$} \epsilon_{ab} \mathbf{s}_m^b \mathbf{s}_m^a + m^2 ) F
\end{equation*}
with gauge fixing functional
\begin{equation*}
F = \hbox{$\frac{1}{2}$} \bigr(
A_\mu^\alpha A^{\mu \alpha} + \xi \epsilon_{ab} C^{\alpha a} C^{\alpha b} + 
\rho (
\epsilon_{cd} C_\mu^{\alpha c} C^{\mu\alpha d} +
\hbox{$\frac{1}{2}$} \sigma \epsilon_{cd} \epsilon_{ef} 
C^{\alpha ce} C^{\alpha df} ) \bigr),
\end{equation*}
$\xi$, $\rho$ and $\sigma$ being the gauge parameters.

For the extended BRST transformations the procedure outlined in (22)--(39) 
yields
\begin{align*}
\mathbf{s}_m^a A_\mu^\alpha &= D_\mu^{\alpha\beta} C^{\beta a} +
C_\mu^{\alpha a},
\\
\mathbf{s}_m^a C^{\alpha b} &= C^{\alpha ab} + \epsilon^{ab} B^\alpha -
\hbox{$\frac{1}{2}$} F^{\alpha\beta\gamma} C^{\beta a} C^{\gamma b},
\\
\mathbf{s}_m^a B^\alpha &= - m^2 C^{\alpha a} + 
B^{\alpha a} +
\hbox{$\frac{1}{2}$} F^{\alpha\beta\gamma} B^\beta C^{\gamma a}
\\
& \quad~ - \hbox{$\frac{1}{2}$} \epsilon_{cd} 
F^{\alpha \beta \gamma} C^{\beta ac} C^{\gamma d} +
\hbox{$\frac{1}{12}$} \epsilon_{cd} 
F^{\alpha \eta \beta} F^{\eta \gamma \delta} 
C^{\gamma a} C^{\delta c} C^{\beta d},
\\
\mathbf{s}_m^a C_\mu^{\alpha b} &= - D_\mu^{\alpha\beta} C^{\beta ab} +
\epsilon^{ab} B_\mu^\alpha - 
F^{\alpha\beta\gamma} C^{\beta a} C_\mu^{\gamma b}, 
\\
\mathbf{s}_m^a B_\mu^\alpha &= - m^2 C_\mu^{\alpha a} - 
D_\mu^{\alpha\beta} B^{\beta a} +
F^{\alpha\beta\gamma} B_\mu^\beta C^{\gamma a} -
\epsilon_{cd} F^{\alpha\beta\gamma} C^{\beta ac} C_\mu^{\gamma d}, 
\\
\mathbf{s}_m^a C^{\alpha bc} &= - \epsilon^{ac} B^{\alpha b} -
\epsilon^{ab} B^{\alpha c} - 
F^{\alpha\beta\gamma} C^{\beta a} C^{\gamma bc},  
\\
\mathbf{s}_m^a B^{\alpha b} &= m^2 C^{\alpha ab} - 
F^{\alpha\beta\gamma} C^{\beta a} B^{\gamma b} +
\hbox{$\frac{1}{2}$} \epsilon_{cd} 
F^{\alpha\beta\gamma} C^{\beta ac} C^{\gamma bd}  
\\
\intertext{and for the corresponding gauge--fixing terms one obtains}
\hbox{$\frac{1}{4}$} \epsilon_{ab} \mathbf{s}_m^b \mathbf{s}_m^a
( A_\mu^\alpha A^{\mu\alpha} ) &= A^{\mu\alpha} ( 
\partial_\mu B^\alpha + B_\mu^\alpha ) +
\hbox{$\frac{1}{2}$} \epsilon_{ab} C^{\mu \alpha b} ( 
\partial_\mu C^{\alpha a} + C_\mu^{\alpha a} )
\\
& \quad~ + \hbox{$\frac{1}{2}$} \epsilon_{ab} ( \partial^\mu C^{\alpha b} ) (
D_\mu^{\alpha \beta} C^{\beta a} + C_\mu^{\alpha a} ),
\\
\hbox{$\frac{1}{4}$} \epsilon_{ab} \mathbf{s}_m^b \mathbf{s}_m^a (
\epsilon_{cd} C^{\alpha c} C^{\alpha d}) &= 
\hbox{$\frac{1}{2}$} m^2 \epsilon_{cd} C^{\alpha c} C^{\alpha d} +
B^\alpha B^\alpha -
\hbox{$\frac{1}{24}$} \epsilon_{ab} \epsilon_{cd}
F^{\eta \alpha \beta} F^{\eta \gamma \delta}
C^{\alpha a} C^{\beta c} C^{\gamma b} C^{\delta d}
\\
& \quad~ - 2 \epsilon_{cd} B^{\alpha c} C^{\alpha d} +
\hbox{$\frac{1}{2}$} \epsilon_{ab} \epsilon_{cd} C^{\alpha ac} C^{\alpha bd} +
\hbox{$\frac{1}{4}$} \epsilon_{ab} \epsilon_{cd} C^{\alpha ac}
F^{\alpha\beta\gamma} C^{\beta b} C^{\gamma d},
\\
\hbox{$\frac{1}{4}$} \epsilon_{ab} \mathbf{s}_m^b \mathbf{s}_m^a (
\epsilon_{cd} C_\mu^{\alpha c} C^{\mu\alpha d}) &=
\hbox{$\frac{1}{2}$} m^2 \epsilon_{cd} C_\mu^{\alpha c} C^{\mu\alpha d} +
B_\mu^\alpha B^{\mu\alpha} +
\hbox{$\frac{1}{2}$} \epsilon_{ab} \epsilon_{cd}
( D_\mu^{\alpha\beta} C^{\beta ac} ) ( D^{\mu\alpha\gamma} C^{\gamma bd} ) 
\\
& \quad~ + 2 \epsilon_{cd} ( D_\mu^{\alpha\beta} B^{\beta c} ) 
C^{\mu\alpha d} +
\epsilon_{ab} \epsilon_{cd} C^{\alpha ac} 
F^{\alpha\beta\gamma} C_\mu^{\beta b} C^{\mu\gamma d}, 
\\
\hbox{$\frac{1}{8}$} \epsilon_{ab} \mathbf{s}_m^b \mathbf{s}_m^a (
\epsilon_{cd} \epsilon_{ef} C_\mu^{\alpha ce} C^{\mu\alpha df}) &=
\hbox{$\frac{1}{2}$} m^2 \epsilon_{cd} \epsilon_{ef} 
C^{\alpha ce} C^{\alpha df} -
\hbox{$\frac{3}{2}$} \epsilon_{cd} B^{\alpha c} B^{\alpha d} + 
\hbox{$\frac{1}{2}$} \epsilon_{cd} \epsilon_{ef} 
F^{\alpha\beta\gamma} B^\alpha C^{\beta ce} C^{\gamma df}
\\
& \quad~ - \hbox{$\frac{1}{4}$} \epsilon_{cd} \epsilon_{ef} 
F^{\alpha\beta\gamma} B^{\alpha c} C^{\beta e} C^{\gamma df} +
\hbox{$\frac{1}{4}$} \epsilon_{ab} \epsilon_{cd} \epsilon_{ef} 
F^{\alpha\beta\gamma} C^{\alpha ad} C^{\beta bf} C^{\gamma ce}.
\end{align*}
The elimination of $B^\alpha$, $B_\mu^\alpha$ and $B^{\alpha a}$ 
by means of gaussian integration provides the gauge--fixing terms 
$\hbox{$\frac{1}{2}$} \xi^{-1} ( \partial^\mu A_\mu^\alpha )^2$ and
$\hbox{$\frac{1}{2}$} \rho^{-1} ( A^{\mu\alpha} )^2$ as well as higher--order 
(anti)ghost interaction terms. In contrast to Ref. \cite{15}, here 
the complete spectrum of (anti)ghosts 
$C^{\alpha a}$, $C_\mu^{\alpha a}$, $C^{\alpha ab}$ and auxiliary fields 
$B^\alpha$, $B_\mu^\alpha$, $B^{\alpha a}$ is produced in a straightforward 
manner. Another advantage consists in the $Sp(2)$--invariant formulation of 
the theory which simplifies all the formulas. 
\bigskip\medskip
\begin{flushleft}
{\large{\bf V. CONCLUDING REMARKS}} 
\end{flushleft}
\bigskip
We have shown that the $osp(1,2)$--symmetric quantization
developed for irreducible massive gauge theories in Ref. \cite{1} can be  
applied also to the case of $L$--stage reducible theories by an appropriate 
generalization of the matrix $(P_+)^{B a}_{A b}$. This formalism 
establishes the well--known fact that mass terms violate gauge independence 
of the $S$--matrix so that after performing BPHZL renormalization,
one has to take the limit of vanishing mass; after that gauge independence 
should be restored. 

Proceeding in the same manner as in Ref. \cite{5} we have built solutions 
of the quantum master equations for massive first--stage reducible 
theories with linearly dependent gauge generators and we found the 
$osp(1,2)$--symmetric realization of the ghost spectrum for general closed 
gauge algebra. Thereby, if compared with the massless case, no extra fields 
had to be introduced. As a consequence of the improved gauge structure 
equation (32) also quartic (anti)ghost terms appear in the extended BRST 
transformations. 

The restriction to theories with closed gauge algebra simplifies the problem 
of finding the full spectrum of (anti)ghosts and auxiliary fields, and the
corresponding symmetry transformations, in so far as it can be done without 
introducing the explicit form of the gauge--fixing terms. Otherwise, for 
theories with open gauge algebra this is no longer possible. The important 
question whether it is possible also in this case to find the full spectrum 
of (anti)ghosts and auxiliary fields together with the corresponding symmetry
transformations in a straightforward manner requires a more detailed 
consideration.  
\bigskip\medskip
\begin{flushleft}
{\large{\bf ACKNOWLEDGEMENT}} 
\end{flushleft}
\bigskip
The work of P. M. Lavrov is partially supported by the Russian Foundation
for Basic Research (RFBR), project 99-02-16617, as well as the grant INTAS, 
96-0308, and grant of Ministry of General and Professional Education of
Russian Federation in field of Basic Natural Sciences. He also thanks Graduate
College ,,Quantum Field Theory'' at Leipzig University for very warm 
hospitality and fruitful cooperation.


\appendix

\bigskip\medskip
\begin{flushleft}
{\large{\bf APPENDIX A. PROOF OF THE IDENTITIES (31) AND (32)}} 
\end{flushleft}
\bigskip
In order to verify the relation (31) we multiply the Jacobi identity (24) 
with $Z^{\alpha_0}_{\beta_1}$; by virtue of 
$R^i_{\alpha_0} Z^{\alpha_0}_{\beta_1} = 0$, this yields 
\begin{align*}
& ( F^{\delta_0}_{\eta_0 \alpha_0} Z^{\alpha_0}_{\beta_1} ) 
F^{\eta_0}_{\beta_0 \gamma_0} +
F^{\delta_0}_{\eta_0 \beta_0}  
( F^{\eta_0}_{\gamma_0 \alpha_0} Z^{\alpha_0}_{\beta_1} ) -
F^{\delta_0}_{\eta_0 \gamma_0}  
( F^{\eta_0}_{\beta_0 \alpha_0} Z^{\alpha_0}_{\beta_1} )
\\
& ~ - R^i_{\beta_0} (
F^{\delta_0}_{\gamma_0 \alpha_0, i} Z^{\alpha_0}_{\beta_1} ) +
R^i_{\gamma_0} (
F^{\delta_0}_{\beta_0 \alpha_0, i} Z^{\alpha_0}_{\beta_1} ) -
Z^{\delta_0}_{\alpha_1} ( 
3 Z^{\alpha_0}_{\beta_1} H^{\alpha_1}_{\alpha_0 \beta_0 \gamma_0} ) = 0.
\end{align*}
After replacing all terms of the form  
$F^{\delta_0}_{\eta_0 \alpha_0} Z^{\alpha_0}_{\beta_1}$ according to 
the relation (28) this gives
\begin{align*}
& Z^{\delta_0}_{\beta_1, i} 
( R^i_{\alpha_0} F^{\alpha_0}_{\beta_0 \gamma_0} ) +
Z^{\delta_0}_{\alpha_1} ( 
G^{\alpha_1}_{\alpha_0 \beta_1} F^{\alpha_0}_{\beta_0 \gamma_0} +
3 Z^{\alpha_0}_{\beta_1} H^{\alpha_1}_{\alpha_0 \beta_0 \gamma_0} )
\\
& ~ + \bigr\{ R^i_{\beta_0} (
F^{\delta_0}_{\gamma_0 \alpha_0, i} Z^{\alpha_0}_{\beta_1} -
F^{\delta_0}_{\alpha_0 \gamma_0} Z^{\alpha_0}_{\beta_1, i} ) -
( F^{\delta_0}_{\alpha_0 \gamma_0} Z^{\alpha_0}_{\alpha_1} )
G^{\alpha_1}_{\beta_0 \beta_1} +
\hbox{antisym}(\beta_0 \leftrightarrow \gamma_0) \bigr\} = 0,
\end{align*}
and, using the same relation once more, 
\begin{align*}
& Z^{\delta_0}_{\beta_1, i} 
( R^i_{\alpha_0} F^{\alpha_0}_{\beta_0 \gamma_0} ) + 
Z^{\delta_0}_{\alpha_1} ( 
G^{\alpha_1}_{\alpha_0 \beta_1} F^{\alpha_0}_{\beta_0 \gamma_0} +
3 Z^{\alpha_0}_{\beta_1} H^{\alpha_1}_{\alpha_0 \beta_0 \gamma_0} )
\\
& ~ + \bigr\{ R^i_{\beta_0} \bigr(
( F^{\delta_0}_{\gamma_0 \alpha_0} Z^{\alpha_0}_{\beta_1} )_{,i} +
Z^{\delta_0}_{\alpha_1, i} G^{\alpha_1}_{\gamma_0 \beta_1} \bigr) +
Z^{\delta_0}_{\alpha_1}  
G^{\alpha_1}_{\beta_0 \gamma_1} G^{\gamma_1}_{\gamma_0 \beta_1} +
\hbox{antisym}(\beta_0 \leftrightarrow \gamma_0) \bigr\} = 0.
\end{align*}
Here, the left--hand side can be rewritten as
\begin{align*}
& Z^{\delta_0}_{\beta_1, i} 
( R^i_{\alpha_0} F^{\alpha_0}_{\beta_0 \gamma_0} ) +
Z^{\delta_0}_{\alpha_1} ( 
G^{\alpha_1}_{\alpha_0 \beta_1} F^{\alpha_0}_{\beta_0 \gamma_0} +
3 Z^{\alpha_0}_{\beta_1} H^{\alpha_1}_{\alpha_0 \beta_0 \gamma_0} )
\\
& ~ + \bigr\{ R^i_{\beta_0} 
( F^{\delta_0}_{\gamma_0 \alpha_0} Z^{\alpha_0}_{\beta_1} +
Z^{\delta_0}_{\alpha_1} G^{\alpha_1}_{\gamma_0 \beta_1} )_{,i} +
Z^{\delta_0}_{\alpha_1} ( 
G^{\alpha_1}_{\gamma_1 \beta_0} G^{\gamma_1}_{\gamma_0 \beta_1} +
R^i_{\gamma_0} G^{\alpha_1}_{\beta_0 \beta_1, i} ) +
\hbox{antisym}(\beta_0 \leftrightarrow \gamma_0) \bigr\} = 0
\end{align*}
and further, once again using relation (28),  
\begin{align}
& Z^{\delta_0}_{\beta_1, i} 
( R^i_{\alpha_0} F^{\alpha_0}_{\beta_0 \gamma_0} ) +
Z^{\delta_0}_{\alpha_1} ( 
G^{\alpha_1}_{\alpha_0 \beta_1} F^{\alpha_0}_{\beta_0 \gamma_0} +
3 Z^{\alpha_0}_{\beta_1} H^{\alpha_1}_{\alpha_0 \beta_0 \gamma_0} )
\nonumber\\
& ~ - \bigr\{ R^i_{\beta_0} 
( Z^{\delta_0}_{\beta_1, j} R^j_{\gamma_0} )_{,i} -
Z^{\delta_0}_{\alpha_1} ( 
G^{\alpha_1}_{\beta_0 \gamma_1} G^{\gamma_1}_{\gamma_0 \beta_1} +
R^i_{\gamma_0} G^{\alpha_1}_{\beta_0 \beta_1, i} ) +
\hbox{antisym}(\beta_0 \leftrightarrow \gamma_0) \bigr\} = 0.
\end{align}
Because of first--stage reducibility and since the gauge algebra (22) 
is closed,
\begin{equation*}
Z^{\delta_0}_{\beta_1, i} 
( R^i_{\alpha_0} F^{\alpha_0}_{\beta_0 \gamma_0} ) =
Z^{\delta_0}_{\beta_1, i} 
( R^j_{\beta_0} R^i_{\gamma_0, j} - R^j_{\gamma_0} R^i_{\beta_0, j} ) =
R^i_{\beta_0} ( Z^{\delta_0}_{\beta_1, j} R^j_{\gamma_0} )_{,i} -
R^i_{\gamma_0} ( Z^{\delta_0}_{\beta_1, j} R^j_{\beta_0} )_{,i},
\end{equation*}
from (40) we get
\begin{equation*}
\bigr(
G^{\alpha_1}_{\beta_0 \gamma_1} G^{\gamma_1}_{\gamma_0 \beta_1} +
R^i_{\beta_0} G^{\alpha_1}_{\gamma_0 \beta_1, i} +
\hbox{antisym}(\beta_0 \leftrightarrow \gamma_0) \bigr) +
G^{\alpha_1}_{\alpha_0 \beta_1} F^{\alpha_0}_{\beta_0 \gamma_0} +
3 Z^{\alpha_0}_{\beta_1} H^{\alpha_1}_{\alpha_0 \beta_0 \gamma_0} = 0
\end{equation*}
which is just the gauge structure relation (31).

In order to prove that the relation (32) is satisfied we have to consider 
the identity
\begin{align*}
\bigr\{& \bigr(
( Z^{\lambda_0}_{\alpha_1} H^{\alpha_1}_{\eta_0 \alpha_0 \beta_0} )
F^{\eta_0}_{\gamma_0 \delta_0} + 
\hbox{cyclic perm} (\alpha_0, \beta_0, \gamma_0) \bigr)
\\
& + 2 R^i_{\delta_0}
( Z^{\lambda_0}_{\alpha_1} H^{\alpha_1}_{\alpha_0 \beta_0 \gamma_0} )_{,i} + 
2 F^{\lambda_0}_{\delta_0 \eta_0} 
( Z^{\eta_0}_{\alpha_1} H^{\alpha_1}_{\alpha_0 \beta_0 \gamma_0} )
\bigr\} + \hbox{antisym}\bigr(\delta_0 \leftrightarrow 
(\alpha_0, \beta_0, \gamma_0) \bigr) \equiv 0,
\end{align*}
which can be verified by a direct calculation by using the Jacobi identity 
(24). Taking into account the relation (28) one obtains 
\begin{align}
Z^{\lambda_0}_{\alpha_1} \bigr\{& \bigr(
H^{\alpha_1}_{\eta_0 \alpha_0 \beta_0} F^{\eta_0}_{\gamma_0 \delta_0} + 
\hbox{cyclic perm} (\alpha_0, \beta_0, \gamma_0) \bigr)
\nonumber\\
& + 2 R^i_{\delta_0} H^{\alpha_1}_{\alpha_0 \beta_0 \gamma_0, i} - 
2 G^{\alpha_1}_{\beta_1 \delta_0} H^{\beta_1}_{\alpha_0 \beta_0 \gamma_0}  
\bigr\} + \hbox{antisym}\bigr(\delta_0 \leftrightarrow 
(\alpha_0, \beta_0, \gamma_0) \bigr) = 0.
\end{align}
After factoring out the zero modes $Z^{\lambda_0}_{\alpha_1}$, and by using
the identity
\begin{align*}
\bigr(&
H^{\alpha_1}_{\eta_0 \alpha_0 \beta_0} F^{\eta_0}_{\gamma_0 \delta_0} + 
\hbox{cyclic perm} (\alpha_0, \beta_0, \gamma_0) \bigr) +
\hbox{antisym}\bigr(\delta_0 \leftrightarrow 
(\alpha_0, \beta_0, \gamma_0) \bigr) 
\\
& \equiv 2 \bigr(
H^{\alpha_1}_{\eta_0 \alpha_0 \beta_0} F^{\eta_0}_{\gamma_0 \delta_0} - 
H^{\alpha_1}_{\eta_0 \delta_0 \alpha_0} F^{\eta_0}_{\beta_0 \gamma_0} + 
\hbox{cyclic perm} (\alpha_0, \beta_0, \gamma_0) \bigr),
\end{align*}
the equation (41) acquires the form 
\begin{align*}
\bigr(& 
H^{\alpha_1}_{\eta_0 \alpha_0 \beta_0} F^{\eta_0}_{\gamma_0 \delta_0} - 
H^{\alpha_1}_{\eta_0 \delta_0 \alpha_0} F^{\eta_0}_{\beta_0 \gamma_0} +
\hbox{cyclic perm} (\alpha_0, \beta_0, \gamma_0) \bigr) 
\\ 
& + \bigr\{ R^i_{\delta_0} H^{\alpha_1}_{\alpha_0 \beta_0 \gamma_0, i} -
G^{\alpha_1}_{\delta_0 \beta_1} H^{\beta_1}_{\alpha_0 \beta_0 \gamma_0} +
\hbox{antisym}\bigr(\delta_0 \leftrightarrow 
(\alpha_0, \beta_0, \gamma_0) \bigr) \bigr\} = 0,
\end{align*}
which is the gauge structure relation (32). Note, that the left--hand 
side of this relation is still a total antisymmetric expression in
$(\alpha_0, \beta_0, \gamma_0, \delta_0)$.
\bigskip\medskip
\begin{flushleft}
{\large{\bf APPENDIX B. COMPONENTWISE NOTATION
\\
OF THE TRANSFORMATIONS (13)}} 
\end{flushleft}
\bigskip
In componentwise notation the extended BRST-- and $Sp(2)$--transformations 
(13) of the antifields reads as follows ($s = 0, \ldots, L$) (the first 
component $D_i$ in $\eta_A$ is put equal to zero):
\begin{align*}
V_m^a \bar{A}_i &= \epsilon^{ab} A_{i b}^*,
\\
V_m^a A_{i b}^* &= m^2 \delta^a_b \bar{A}_i,
\\
V_m^a \bar{B}_{\alpha_s|a_1 \cdots a_s} &= \epsilon^{ab}
B_{\alpha_s b|a_1 \cdots a_s}^*,
\\
V_m^a E_{\alpha_s|a_1 \cdots a_s} &= m^2 \epsilon^{ab} ( 
s B_{\alpha_s b|a_1 \cdots a_s}^* -
\sum_{r = 1}^s B_{\alpha_s a_r|a_1 \cdots a_{r - 1} b 
a_{r + 1} \cdots a_s}^* ),
\\
V_m^a B_{\alpha_s b|a_1 \cdots a_s}^* &= m^2 ( 
\delta^a_b \bar{B}_{\alpha_s|a_1 \cdots a_s} + 
\sum_{r = 1}^s \delta^a_{a_r}
\bar{B}_{\alpha_s|a_1 \cdots a_{r - 1} b a_{r + 1} \cdots a_s} ) -
\delta^a_b E_{\alpha_s|a_1 \cdots a_s},
\\
V_m^a \bar{C}_{\alpha_s|a_0 \cdots a_s} &= \epsilon^{ab}
C_{\alpha_s b|a_0 \cdots a_s}^*,
\\
V_m^a F_{\alpha_s|a_0 \cdots a_s} &= m^2 \epsilon^{ab} ( 
(s + 1) C_{\alpha_s b|a_0 \cdots a_s}^* -
\sum_{r = 0}^s C_{\alpha_s a_r|a_0 \cdots a_{r - 1} b
a_{r + 1} \cdots a_s}^* ),
\\
V_m^a C_{\alpha_s b|a_0 \cdots a_s}^* &= m^2 ( 
\delta^a_b \bar{C}_{\alpha_s|a_0 \cdots a_s} + 
\sum_{r = 0}^s \delta^a_{a_r}
\bar{C}_{\alpha_s|a_0 \cdots a_{r - 1} b a_{r + 1} \cdots a_s} ) -
\delta^a_b F_{\alpha_s|a_0 \cdots a_s}
\\
\intertext{and}
V_\alpha \bar{A}_i &= 0,
\\
V_\alpha A_{i b}^* &= A_{i c}^* (\sigma_\alpha)^c_{~b}, 
\\
V_\alpha \bar{B}_{\alpha_s|a_1 \cdots a_s} &= \sum_{r = 1}^s
\bar{B}_{\alpha_s|a_1 \cdots a_{r - 1} c a_{r + 1} \cdots a_s} 
(\sigma_\alpha)^c_{~a_r},
\\
V_\alpha B_{\alpha_s b|a_1 \cdots a_s}^* &= 
B_{\alpha_s c|a_1 \cdots a_s}^* (\sigma_\alpha)^c_{~b} + \sum_{r = 1}^s
B_{\alpha_s b|a_1 \cdots a_{r - 1} c a_{r + 1} \cdots a_s}^* 
(\sigma_\alpha)^c_{~a_r},
\\
V_\alpha \bar{C}_{\alpha_s|a_0 \cdots a_s} &= \sum_{r = 0}^s
\bar{C}_{\alpha_s|a_0 \cdots a_{r - 1} c a_{r + 1} \cdots a_s} 
(\sigma_\alpha)^c_{~a_r},
\\
V_\alpha C_{\alpha_s b|a_0 \cdots a_s}^* &= 
C_{\alpha_s c|a_0 \cdots a_s}^* (\sigma_\alpha)^c_{~b} + \sum_{r = 0}^s
C_{\alpha_s b|a_0 \cdots a_{r - 1} c a_{r + 1} \cdots a_s}^* 
(\sigma_\alpha)^c_{~a_r},
\\
V_\alpha E_{\alpha_s|a_1 \cdots a_s} &= \sum_{r = 1}^s
E_{\alpha_s|a_1 \cdots a_{r - 1} c a_{r + 1} \cdots a_s} 
(\sigma_\alpha)^c_{~a_r},
\\
V_\alpha F_{\alpha_s|a_0 \cdots a_s} &= \sum_{r = 0}^s
F_{\alpha_s|a_0 \cdots a_{r - 1} c a_{r + 1} \cdots a_s} 
(\sigma_\alpha)^c_{~a_r}.
\end{align*}
where the additional sources $E_{\alpha_s|a_1 \cdots a_s}$ and
$F_{\alpha_s|a_0 \cdots a_s}$ have to be introduced in order to satisfy 
the $osp(1,2)$--superalgebra
\begin{equation*}
[ V_\alpha, V_\beta ] = \epsilon_{\alpha\beta}^{~~~\!\gamma} V_\gamma,
\qquad
[ V_\alpha, V_m^a ] = V_m^b (\sigma_\alpha)_b^{~a},
\qquad
\{ V_m^a, V_m^b \} = - m^2 (\sigma^\alpha)^{ab} V_\alpha.
\end{equation*}
Let us emphasize that expressing this algebra through operator identities
is a stronger restriction than satisfying this algebra by the help of
(anti)BRST transformations which can be realized without
introducing $E_{\alpha_s|a_1 \cdots a_s}$ and
$F_{\alpha_s|a_0 \cdots a_s}$, namely by choosing also the antifields
$B_{\alpha_s b|a_1 \cdots a_s}^*$ and $C_{\alpha_s b|a_0 \cdots a_s}^*$
as irreducible representations, i.e. totally symmetric with respect to all
$Sp(2)$--indicies, $B_{\alpha_s b|a_1 \cdots a_s}^* = 
B_{\alpha_s a_r|a_1 \cdots a_{r - 1} b a_{r + 1} \cdots a_s}^*$ 
for $r = 1, \ldots, s$ and 
$C_{\alpha_s b|a_0 \cdots a_s}^* = 
C_{\alpha_s a_r|a_0 \cdots a_{r - 1} b a_{r + 1} \cdots a_s}^*$ 
for $r = 0, \ldots, s$ ($s = 0, \ldots, L$).

Let us also write down the componentwise notation of the operators 
$V_m^a$, $V_\alpha$ and $\Delta^a$, $\Delta_\alpha$, Eqs. (10)--(12).
They are given by
\begin{align*}
V_m^a &= \epsilon^{ab} A_{i b}^* \frac{\delta}{\delta \bar{A}_i} +
m^2 \bar{A}_i \frac{\delta}{\delta A_{i a}^*} + \sum_{s = 0}^L \Bigr\{
\epsilon^{ab} B_{\alpha_s b|a_1 \cdots a_s}^*
\frac{\delta}{\delta \bar{B}_{\alpha_s|a_1 \cdots a_s}}
\\
& \quad~ +
m^2 ( \delta^a_b \bar{B}_{\alpha_s|a_1 \cdots a_s} +
\sum_{r = 1}^s \delta^a_{a_r}
\bar{B}_{\alpha_s|a_1 \cdots a_{r - 1} b a_{r + 1} \cdots a_s} )
\frac{\delta}{\delta B_{\alpha_s b|a_1 \cdots a_s}^*} 
\\
& \quad~ - E_{\alpha_s|a_1 \cdots a_s}
\frac{\delta}{\delta B_{\alpha_s a|a_1 \cdots a_s}^*} 
+ m^2 \epsilon^{ab} \sum_{r = 1}^s (
B_{\alpha_s b|a_1 \cdots a_s}^* -
B_{\alpha_s a_r|a_1 \cdots a_{r - 1} b a_{r + 1} \cdots a_s}^* )
\frac{\delta}{\delta E_{\alpha_s|a_1 \cdots a_s}}
\\
& \quad~ + \epsilon^{ab} C_{\alpha_s b|a_0 \cdots a_s}^*
\frac{\delta}{\delta \bar{C}_{\alpha_s|a_0 \cdots a_s}} +
m^2 ( \delta^a_b \bar{C}_{\alpha_s|a_0 \cdots a_s} +
\sum_{r = 0}^s \delta^a_{a_r}
\bar{C}_{\alpha_s|a_0 \cdots a_{r - 1} b a_{r + 1} \cdots a_s} )
\frac{\delta}{\delta C_{\alpha_s b|a_0 \cdots a_s}^*} 
\\
& \quad~ - F_{\alpha_s|a_0 \cdots a_s}
\frac{\delta}{\delta C_{\alpha_s a|a_0 \cdots a_s}^*} 
+ m^2 \epsilon^{ab} \sum_{r = 0}^s (
C_{\alpha_s b|a_0 \cdots a_s}^* -
C_{\alpha_s a_r|a_0 \cdots a_{r - 1} b a_{r + 1} \cdots a_s}^* )
\frac{\delta}{\delta F_{\alpha_s|a_0 \cdots a_s}} \Bigr\}
\\
V_\alpha &= A_{i c}^* (\sigma_\alpha)^c_{~b}
\frac{\delta}{\delta A_{i b}^*} + \sum_{s = 0}^L \Bigr\{
\sum_{r = 1}^s 
\bar{B}_{\alpha_s|a_1 \cdots a_{r - 1} c a_{r + 1} \cdots a_s} 
(\sigma_\alpha)^c_{~a_r}
\frac{\delta}{\delta \bar{B}_{\alpha_s|a_1 \cdots a_s}}
\\
&\quad + \sum_{r = 1}^s 
E_{\alpha_s|a_1 \cdots a_{r - 1} c a_{r + 1} \cdots a_s} 
(\sigma_\alpha)^c_{~a_r}
\frac{\delta}{\delta E_{\alpha_s|a_1 \cdots a_s}}  
\\
& \quad~ + \bigr(
B_{\alpha_s c|a_1 \cdots a_s}^* (\sigma_\alpha)^c_{~b} + \sum_{r = 1}^s
B_{\alpha_s b|a_1 \cdots a_{r - 1} c a_{r + 1} \cdots a_s}^* 
(\sigma_\alpha)^c_{~a_r} \bigr)
\frac{\delta}{\delta B_{\alpha_s b|a_1 \cdots a_s}^*}
\\
&\quad + \sum_{r = 0}^s 
\bar{C}_{\alpha_s|a_0 \cdots a_{r - 1} c a_{r + 1} \cdots a_s} 
(\sigma_\alpha)^c_{~a_r}
\frac{\delta}{\delta \bar{C}_{\alpha_s|a_0 \cdots a_s}}
\\
&\quad~ + \sum_{r = 0}^s 
F_{\alpha_s|a_0 \cdots a_{r - 1} c a_{r + 1} \cdots a_s} 
(\sigma_\alpha)^c_{~a_r}
\frac{\delta}{\delta F_{\alpha_s|a_0 \cdots a_s}}  
\\
& \quad~ + \bigr(
C_{\alpha_s c|a_0 \cdots a_s}^* (\sigma_\alpha)^c_{~b} + \sum_{r = 0}^s
C_{\alpha_s b|a_0 \cdots a_{r - 1} c a_{r + 1} \cdots a_s}^* 
(\sigma_\alpha)^c_{~a_r} \bigr)
\frac{\delta}{\delta C_{\alpha_s b|a_0 \cdots a_s}^*} \Bigr\}
\\
\intertext{and}
\Delta^a &= (-1)^{\epsilon_i}
\frac{\delta_L}{\delta A^i} \frac{\delta}{\delta A_{i a}^*} + 
\sum_{s = 0}^L \Bigr\{ 
(-1)^{\epsilon_{\alpha_s}} \sum_{r = 1}^s 
\frac{\delta_L}{\delta B^{\alpha_s|a_1 \cdots a_s}} 
\frac{\delta}{\delta B_{\alpha_s a|a_1 \cdots a_s}^*}
\\
& \quad\qquad\qquad\qquad\qquad\qquad + 
(-1)^{\epsilon_{\alpha_s} + 1} \sum_{r = 0}^s 
\frac{\delta_L}{\delta C^{\alpha_s|a_0 \cdots a_s}} 
\frac{\delta}{\delta C_{\alpha_s a|a_0 \cdots a_s}^*} \Bigr\}
\\
\Delta_\alpha &= \sum_{s = 0}^L \Bigr\{
(-1)^{\epsilon_{\alpha_s}} \sum_{r = 1}^s (\sigma_\alpha)_c^{~a_r} 
\frac{\delta_L}{\delta B^{\alpha_s|a_1 \cdots a_s}} 
\frac{\delta}{\delta E_{\alpha_s|a_1 \cdots a_{r - 1} c a_{r + 1} \cdots a_s}}
\\
& \qquad\quad~ + 
(-1)^{\epsilon_{\alpha_s} + 1} \sum_{r = 0}^s (\sigma_\alpha)_c^{~a_r}
\frac{\delta_L}{\delta C^{\alpha_s|a_0 \cdots a_s}} 
\frac{\delta}{\delta F_{\alpha_s|a_0 \cdots a_{r - 1} c a_{r + 1} \cdots a_s}} 
\Bigr\}.
\end{align*}


\end{document}